\documentclass[longauth,breaklinks,pdfnewwindow=true,colorlinks=true,citecolor=blue]{aa}  

\usepackage{threeparttable}
\usepackage{graphicx}
\usepackage{gensymb}
\usepackage{multirow}
\usepackage[T1]{fontenc}
\usepackage{amssymb}
\usepackage{natbib}
\usepackage{tabularx}
\usepackage[utf8]{inputenc}
\usepackage[english]{babel}
\usepackage{txfonts}
\usepackage{ulem}
\usepackage{xcolor}
\usepackage{soul}
\usepackage{orcidlink}  
\usepackage{hyperref}
\usepackage[switch]{lineno}

\newcommand{\orcid}[1]{\unskip\protect\href{https://orcid.org/#1}{\protect\includegraphics[width=8pt,clip]{logo_orcid}}}

\newcommand{\beam}{$\theta_{\mbox{\scriptsize maj}}\times\theta_{\mbox{\scriptsize min}}$}

\begin{document}

   \title{FAUST}
   \subtitle{XXXI. Grain properties and variability of three sources in GSS 30}

   \author{Qiancheng Yang
          \inst{1}\orcidlink{0009-0001-5048-3034}
          \and
          Hauyu Baobab Liu
          \inst{2,3}\orcidlink{0000-0003-2300-2626}
          \and
          Siyi Feng 
          \inst{1}\thanks{Corresponding author: syfeng@xmu.edu.cn}\orcidlink{0000-0002-4707-8409}
          \and
          C. J. Chandler
          \inst{4}\orcidlink{0000-0002-7570-5596}
          \and
          F. Fontani
          \inst{5,6,7}\orcidlink{0000-0003-0348-3418}
          \and
          N. Sakai
          \inst{8}\orcidlink{0000-0002-3297-4497}
          \and
          Y. Oya
          \inst{9}\orcidlink{0000-0002-0197-8751}
          \and
          T. Hanawa
          \inst{10}\orcidlink{0000-0002-7538-581X}
          \and
          C. Codella
          \inst{5,11}\orcidlink{0000-0003-1514-3074}
          \and
          C. Ceccarelli
          \inst{11}\orcidlink{0000-0001-9664-6292}
          \and
          I. Jim{\'e}nez-Serra
          \inst{12}\orcidlink{0000-0003-4493-8714}
          \and
          L. Cacciapuoti
          \inst{13}\orcidlink{0000-0001-8266-0894}
          \and
          E. Bianchi
          \inst{5}\orcidlink{0000-0001-9249-7082}
          \and
          M. Bouvier
          \inst{14}\orcidlink{0000-0003-0167-0746}
          \and
          M. De Simone
          \inst{5,15}\orcidlink{0000-0001-5659-0140}
          \and
          A. Isella
          \inst{16}\orcidlink{0000-0001-8061-2207}\
          \and
          D. Johnstone
          \inst{17,18}\orcidlink{0000-0002-6773-459X}
          \and
          L. Loinard
          \inst{19,20,21}\orcidlink{0000-0002-5635-3345}
          \and
          G. Sabatini
          \inst{5}\orcidlink{0000-0002-6428-9806}
          \and
          Y. Shirley
          \inst{22}\orcidlink{0000-0002-0133-8973}
          \and
          Y. Aikawa
          \inst{23}\orcidlink{0000-0003-3283-6884}
          \and
          M. J. Maureira
          \inst{24}\orcidlink{0000-0002-7026-8163}
          \and
          P. Caselli
          \inst{24}\orcidlink{0000-0003-1481-7911}
          \and
          F. Menard
          \inst{11}\orcidlink{0000-0002-1637-7393}
          \and
          N. Balucani
          \inst{25}\orcidlink{0000-0001-5121-5683}
          \and
          E. Caux
          \inst{26}\orcidlink{0000-0002-4463-6663}
          \and
          S. Charnley
          \inst{27}\orcidlink{0000-0001-6752-5109}
          \and
          N. Cuello
          \inst{11}\orcidlink{0000-0003-3713-8073}
          \and
          F. Dulieu
          \inst{28}\orcidlink{0000-0001-6981-0421}
          \and
          L. Evans
          \inst{29}\orcidlink{0009-0006-1929-3896}
          \and
          D. Fedele
          \inst{5}\orcidlink{0000-0001-6156-0034}
          \and
          T. Hama
          \inst{30,31}\orcidlink{0000-0002-4991-4044}
          \and
          E. Herbst
          \inst{32}\orcidlink{0000-0002-4649-2536}
          \and
          T. Hirota
          \inst{33,34}\orcidlink{0000-0003-1659-095X}
          \and
          B. Lefloch
          \inst{35}\orcidlink{0000-0002-9397-3826}
          \and
          A. L{\'o}pez-Sepulcre
          \inst{11,36}\orcidlink{0000-0002-6729-3640}
          \and
          L. T. Maud
          \inst{15}\orcidlink{0000-0002-7675-3565}
          \and
          A. Miotello
          \inst{15}\orcidlink{0000-0002-7997-2528}
          \and
          G. Moellenbrock
          \inst{4}\orcidlink{0000-0002-3296-8134}
          \and
          H. Nomura
          \inst{37,34}\orcidlink{0000-0002-7058-7682}
          \and
          Y. Oba
          \inst{38}\orcidlink{0000-0002-6852-3604}
          \and
          S. Ohashi
          \inst{8,37}\orcidlink{0000-0002-9661-7958}
          \and
          Y. Okoda
          \inst{8,17}\orcidlink{0000-0003-3655-5270}
          \and
          J. Pineda
          \inst{24}\orcidlink{0000-0002-3972-1978}
          \and
          L. Podio
          \inst{5}\orcidlink{0000-0003-2733-5372}
          \and
          A. Rimola
          \inst{39}\orcidlink{0000-0002-9637-4554}
          \and
          T. Sakai
          \inst{40}\orcidlink{0000-0003-4521-7492}
          \and
          D. Segura-Cox
          \inst{41}\orcidlink{0000-0003-3172-6763}
          \and
          B. Svoboda
          \inst{4}\orcidlink{0000-0002-8502-6431}
          \and
          L. Testi
          \inst{5,42}\orcidlink{0000-0003-1859-3070}
          \and
          C. Vastel
          \inst{26}\orcidlink{0000-0001-8211-6469}
          \and
          S. Viti
          \inst{14,43}\orcidlink{0000-0001-8504-8844}
          \and
          N. Watanabe
          \inst{39}\orcidlink{0000-0001-8408-2872}
          \and
          Y. Watanabe
          \inst{44}\orcidlink{0000-0002-9668-3592}
          \and
          Y. Zhang
          \inst{45}\orcidlink{0000-0001-7511-0034}
          \and
          Z. E. Zhang
          \inst{46}
          \and
          F. O. Alves
          \inst{36}\orcidlink{0000-0002-7945-064X}
          \and
          S. Yamamoto
          \inst{47}\orcidlink{0000-0002-9865-0970} 
          }

\titlerunning{FAUST XXXI.  Grain properties and variability of three sources in GSS 30}
\authorrunning{Yang et al.}

   \institute{
             Department of Astronomy, Xiamen University, Zengcuoan West Road, Xiamen, 361005
        \and
             Department of Physics, National Sun Yat-Sen University, No. 70, Lien-Hai Road, Kaohsiung City 80424
        \and
             Center of Astronomy and Gravitation, National Taiwan Normal University, Taipei 116
        \and
             National Radio Astronomy Observatory, PO Box O, Socorro, NM 87801, USA
        \and
             Istituto Nazionale di Astrofisica (INAF), Osservatorio Astrofisico di Arcetri, Largo E. Fermi 5, 50125 Firenze, Italy
        \and
             Max-Planck-Institut f{\"u}r extraterrestrische Physik (MPE), Gie$\beta$enbachstr. 1, D-85741 Garching, Germany
        \and
             LUX, Observatoire de Paris, PSL Research University, CNRS, Sorbonne Universit\'{e}s, 92190 Paris, France
        \and
             RIKEN Cluster for Pioneering Research, 2-1, Hirosawa, Wako-shi, Saitama 351-0198, Japan
        \and
             Center for Gravitational Physics, Yukawa Institute for Theoretical Physics, Kyoto University, Oiwake-cho, Kitashirakawa, Sakyo-ku, Kyoto-shi, Kyoto-fu 606-8502, Japan
        \and
             Center for Frontier Science, Chiba University, 1-33 Yayoi-cho, Inage-ku, Chiba 263-8522, Japan
        \and
             Univ. Grenoble Alpes, CNRS, IPAG, 38000 Grenoble, France
        \and
             Centro de Astrobiolog{\'i}a (CAB), INTA-CSIC, Ctra. de Torrej{\'o}n a Ajalvir, km 4, 28850, Torrej{\'o}n de Ardoz, Spain
        \and
             European Southern Observatory, Alonso de Cordova 3107, Vitacura, Region Metropolitana de Santiago, Chile
        \and
             Leiden Observatory, Leiden University, PO Box 9513, 2300 RA Leiden, The Netherlands
        \and
             European Southern Observatory, Karl-Schwarzschild Str. 2, 85748 Garching bei M{\"u}nchen, Germany
        \and
             Department of Physics and Astronomy, Rice University, 6100 Main Street, MS-108, Houston, TX 77005, USA
        \and
             NRC Herzberg Astronomy and Astrophysics, 5071 West Saanich Road, Victoria, BC, V9E 2E7, Canada
        \and
             Department of Physics and Astronomy, University of Victoria, Victoria, BC V8P 5C2, Canada
        \and
             Instituto de Radioastronom{\'i}a y Astrofísica , Universidad Nacional Aut{\'o}noma de M{\'e}xico, Apartado Postal 3-72, Morelia 58090, Michoac{\'a}n, Mexico
        \and
             Black Hole Initiative at Harvard University, 20 Garden Street, Cambridge, MA 02138, USA
        \and
             David Rockefeller Center for Latin American Studies, Harvard University, 1730 Cambridge Street, Cambridge, MA 02138, USA
        \and
             Steward Observatory, 933 N Cherry Ave., Tucson, AZ 85721, USA
        \and
             Department of Astronomy, The University of Tokyo, 7-3-1 Hongo, Bunkyo-ku, Tokyo 113-0033, Japan
        \and
             Center for Astrochemical Studies, Max-Planck-Institut f{\"u}r extraterrestrische Physik (MPE), Gie$\beta$enbachstr. 1, 85741 Garching, Germany
        \and
             Department of Chemistry, Biology, and Biotechnology, The University of Perugia, Via Elce di Sotto 8, 06123 Perugia, Italy
        \and
             IRAP, Univ. de Toulouse, CNRS, CNES, UPS, Toulouse, France
        \and
             Astrochemistry Laboratory, Code 691, NASA Goddard Space Flight Center, 8800 Greenbelt Road, Greenbelt, MD 20771, USA
        \and
             CY Cergy Paris Universit{\'e}, Sorbonne Universit{\'e}, Observatoire de Paris, PSL University, CNRS, LERMA, 95000, Cergy, France
        \and
             School of Physics and Astronomy, University of Leeds, Leeds LS29JT, UK
        \and
             Komaba Institute for Science, The University of Tokyo, 3-8-1 Komaba, Meguro, Tokyo 153-8902, Japan
        \and
             Department of Basic Science, The University of Tokyo, 3-8-1 Komaba, Meguro, Tokyo 153-8902, Japan
        \and
             Department of Chemistry, University of Virginia, McCormick Road, PO Box 400319, Charlottesville, VA 22904, USA
        \and
             National Astronomical Observatory of Japan, 2-12 Hoshigaoka, Mizusawa, Oshu, Iwate 023-0861, Japan
        \and
             SOKENDAI (The Graduate University for Advanced Studies), 2-21-1 Osawa, Mitaka, Tokyo 181-8588, Japan
        \and
             Universit{\'e} de Bordeaux – CNRS Laboratoire d’Astrophysique de Bordeaux, 33600 Pessac, France
        \and
             Institut de Radioastronomie Millim{\'e}trique, 38406 Saint-Martin d’H\`{e}res, France
        \and
             National Astronomical Observatory of Japan, 2-21-1 Osawa, Mitaka, Tokyo 181-8588, Japan
        \and
             Institute of Low Temperature Science, Hokkaido University, N19W8, Kita-ku, Sapporo, Hokkaido 060-0819, Japan
        \and
             Departament de Qu{\'i}mica, Universitat Aut\`{o}noma de Barcelona, 08193 Bellaterra, Spain
        \and
             Graduate School of Informatics and Engineering, The University of Electro-Communications, Chofu, Tokyo 182-8585, Japan
        \and
             Department of Astronomy, The University of Texas at Austin, 2515 Speedway, Austin, Texas 78712, USA
        \and
             Alma Mater Studiorum – Universit\`{a} di Bologna, Dipartimento di Fisica e Astronomia ``Augusto Righi'', Via Gobetti 93/2, I-40129, Bologna, Italy
        \and
             Transdisciplinary Research Area (TRA) `Matter'/Argelander-Institut für Astronomie, University of Bonn
        \and
             Materials Science and Engineering, College of Engineering, Shibaura Institute of Technology, 3-7-5 Toyosu, Koto-ku, Tokyo 135-8548, Japan
        \and
             Department of Astronomy, Shanghai Jiao Tong University, 800 Dongchuan Rd., Minhang, Shanghai 200240
        \and
             Star and Planet Formation Laboratory, RIKEN Cluster for Pioneering Research, Wako, Saitama 351-0198, Japan
        \and
             SOKENDAI (The Graduate University for Advanced Studies), Shonan Village, Hayama, Kanagawa 240-0193, Japan
             }

   \date{Received XXX; accepted YYY}

\abstract
   {} 
   {
   To advance our understanding of dust properties in Class 0/I young stellar objects (YSOs), it is essential to resolve and characterize their structures at multiple wavelengths, and to investigate how both grain growth and environmental factors shape their spectral properties.}
   {We present 0\farcs5 angular resolution ALMA observations of the GSS 30 complex at 1.2–3.0\,mm, collected as part of the FAUST large program, with a linear resolution of 69 au. We analyze the dust continuum emission and perform model fitting to constrain the dust mass, the disk structure.
   }
   {For IRS3, the spectral index increases radially from 2.0 at the center to 2.5 at the disk edge. In contrast, it decreases from 2.0 to as low as 1.6–1.8 along the outflow direction. An asymmetric distribution shows $\alpha$ as low as 1.6 toward the northeastern blue-shifted lobe, possibly due to cold outer envelope dust obscuring warmer inner regions. Spectral energy distribution (SED) fitting indicates maximum dust grain sizes of tens of microns and a dust mass of 650-1510 M$_\oplus$. IRS1 is connected with an extended structure to the northeast. It may be an outflow–disk complex or a trailing structure associated with a circumbinary disk, although we cannot rule out the possibility that it could also be a gas streamer accreting to IRS1. Toward the center of IRS1, the spectral index is $<$0.8, which is consistent with marginally optically thick free-free emission. IRS2 exhibits brightness variations over 420 seconds of observation, with multi-epoch comparison suggesting a flare lasting tens of minutes, indicating magnetic activity in the central star.}
   {This study presents a detailed millimeter multi-wavelength analysis of the GSS 30 IRS3 and IRS1 systems, highlighting the crucial role of environmental factors-specifically the dust obscuration and the connecting streamer structures-in shaping the observed morphology of these early-stage disks. We also find that the IRS2 exhibits millimeter variability which may originate from magnetic flares at the surface of the protostar. 
   }

   \maketitle

%________________________________________________________________

\section{Introduction}\label{sec:introduction}

Despite significant progress in understanding protoplanetary disks by using interferometers, many questions with respect to their evolution remain unanswered.
A prominent issue is how the solids grow from submicron interstellar medium (ISM) sizes to planetary scales. Atacama Large Millimeter/submillimeter Array (ALMA) large-scale surveys have revealed abundant substructures, such as rings, gaps \citep{2015ApJ...808L...3A} and spiral arms \citep{2012ApJ...748L..22M,2014ApJ...785L..12C,2022NatAs...6..331D,2022NatAs...6..837L} in Class II protoplanetary disks (e.g., DSHARP, \citealt{2018ApJ...869L..41A}; MAPS, \citealt{2021ApJS..257....1O}; AGE-PRO, \citealt{2024ApJ...974..102S}). One of the interpretations of these features is the presence, or ongoing formation, of planets massive enough to carve out structures in the disk, implying the existence of kilometer-sized planetesimals or even larger bodies \citep{2023ASPC..534..423B}. Recent observations show these substructures exist even in earlier Class 0/I protostellar disks, compressing the solids growth timescale to within 1 Myr 
(\citealt{2018ApJ...857...18S,2020Natur.586..228S,2020ApJ...895L...2N}; FAUST, \citealt{2024A&A...689L...5M}; eDISK, \citealt{2023ApJ...951....8O}).

Inferring dust properties from (sub)millimeter observations is challenging due to various physical and observational uncertainties. A widely used method is to estimate the wavelength dependent dust opacity via measuring the spectral indices of thermal emission in the Rayleigh–Jeans regime. The dust opacity index ($\beta$) provides insight into grain growth and can be indirectly inferred from the observed spectral index ($\alpha$ = $\beta$+2). In the diffuse ISM, $\beta$ is typically around 1.7 (e.g., \citealt{1982ApJ...252..589S,2014A&A...564A..45P,2016ApJ...828...32L}), while lower values ($\beta$ $\lesssim$ 1) observed in T Tauri disks suggest grain growth to millimeter sizes (e.g., \citealt{1990AJ.....99..924B,1991ApJ...381..250B,2006ApJ...636.1114D,2010A&A...521A..66R,2010A&A...512A..15R,2017A&A...604A..52B,2021MNRAS.506.2804T,2021MNRAS.506.5117T}). Recent ALMA observations of Class 0/I systems revealed low $\beta$ values in both disks and envelopes, pointing to possible early grain growth within the natal core (e.g., \citealt{2019A&A...632A...5G,2020ApJ...890..130T,2020A&A...640A..19T,2021A&A...653A.117B,2023A&A...676A...4C,2025A&A...700A.188C,2025A&A...698L..16S}). 
While estimates of $\beta$ from spectral energy distributions (SEDs) fitting are widely used to diagnose grain growth, some sources show unusually low spectral indices ($\alpha$ < 2) that cannot be fully explained by grain growth alone (e.g., \citealt{2007ApJ...659..479J,2014A&A...567A..32M,2017ApJ...840...72L,2018A&A...612A..54L,2019A&A...623A.147A}), possibly involving free–free emission contamination \citep{1986ApJ...304..713R}, line-of-sight obscuration \citep{2017ApJ...840...72L,2018ApJ...868...39G,2019ApJ...884...97L}, or dust self-scattering \citep{2019ApJ...877L..22L}.

Moreover, due to the limited angular resolution and sensitivity of earlier instruments, many complex physical processes in protostellar systems remained poorly characterized. 
Only recently, thanks to the improved angular resolution and sensitivity of modern interferometry, it has been understood that the environmental effects are far more complicated than expected. 
In particular, substantial progress has been made in understanding the roles of magnetic fields \citep[e.g., ][]{2019FrASS...6....3H,2019FrASS...6...15P,2020ApJ...904..168B,2022FrASS...9.3556L}, turbulence \citep[e.g., ][]{2015A&A...581A.119B,2022ApJ...926..165L,2023ApJ...949..109L,2024ApJ...969...70F}, and protostellar outflows \citep[e.g., ][]{2016ARA&A..54..491B,2016Natur.540..406B,2017A&A...603L...3A,2017NatAs...1E.152L,2019ApJ...874..104K,2024A&A...684L..12S,2024ApJ...961...90C,2025A&A...698L..16S}. Other environmental features, such as streamer-like structures connecting envelopes to molecular clouds \citep[e.g., ][]{2020NatAs...4.1158P,2022A&A...658A..53M,2022A&A...667A..12V,2023A&A...677A..92V,2024A&A...682A..61C,2025A&A...701A.165G} and millimeter flares \citep[e.g., ][]{2023MNRAS.522...56V,2025ApJ...989...11M}, have also been observed with ALMA. Unfortunately, these many effects remain insufficiently explored.

The low-mass star-forming region GSS 30, nestled within the L1688 region of the Ophiuchus molecular cloud at a distance of 138.4 ± 2.6 pc (\citealt{1973ApJ...184L..53G, 1978ApJ...224..453E, 1993AJ....105..271W, 2008AN....329...10M, 2018ApJ...869L..33O}) holds particular intrigue. Over a spatial extent of 2000 au, this contains three infrared sources: two low-mass protostars– IRS1 in the south and IRS3 (also known as LFAM1 in \citealt{1991ApJ...379..683L}) in the north, alongside a T-Tauri star IRS2, located to the east. The region is also associated with a bipolar reflection nebula. IRS2 was detected at 6 cm (\citealt{1959SvA.....3..434D, 1991ApJ...379..683L, 1993AJ....105..271W}), and its association with the same nebula was confirmed by VLBA \citep{2017ApJ...834..141O,2018ApJ...869L..33O}. 
The SED fitting across wavelengths from microns to centimeters suggest that the IRS1-IRS3 system is in a young evolutionary stage (e.g., \citealt{1994ApJ...420..837A, 1995ApJ...445..377C, 2001A&A...372..173B, 2009ApJ...692..973E, 2013ApJ...770..123G}). 
The deconvolved \textit{Herschel}/PACS spectra distinguish this system, IRS1 was classified as a Class I source, with a bolometric luminosity of 8.5 L$_\sun$ and a bolometric temperature of 193 K \citep{2015ApJS..217....6J}, whereas IRS3 is classified as a younger Class 0 object, characterized by a lower bolometric luminosity of 1.7 L$_\sun$) and bolometric temperature of 50 K\citep{2023ApJ...951....8O}.
Moreover, line emission indicates the presence of outflows in this region (\citealt{2015MNRAS.447.1996W}). Previous observations with high angular resolution radio interferometers, such as the Owens Valley Radio Observatory Millimeter Array (OVRO) (\citealt{1997ApJ...475..713Z}), the Submillimeter Array (SMA) (e.g., \citealt{2009A&A...507..861J}), and the ALMA (e.g., \citealt{2018ApJ...869..158F, 2019ApJS..245....2S,2024A&A...690A..46S}), have successfully resolved IRS1 and IRS3 as distinct sources. 
These studies indirectly suggest the presence of disk-envelope systems around both protostars (\citealt{2011A&A...533A.112H, 2018ApJ...869..158F, 2019ApJS..245....2S, 2020A&A...636A..26H}).

In this paper, we present observations of GSS 30 at $\sim$0\farcs5 angular resolution  in the continuum as part of the ALMA large project “Fifty AU STudy of the chemistry in the disk/envelope system of Solar-like protostars” (FAUST)\footnote{\url{http://faust-alma.riken.jp}} using ALMA band 3 and band 6 \citep[PI: S. Yamamoto,][]{2021FrASS...8..227C}. We provide an overview of the data quality obtained from ALMA observations in Section \ref{sec:observations}. Then, we present the continuum maps and investigate potential continuum variability of IRS2 in Section \ref{sec:result}. In Section \ref{sec:discussion}, we discuss the results of our analysis and summarize our main conclusions in Section \ref{sec:conclusions}.

\section{Observations}\label{sec:observations}

GSS 30 IRS3 was observed between 2018 November and 2019 September, assuming as reference coordinate $\alpha_{2000} = \rm 16h26m21s.72$ and $\delta_{2000} = \rm -24\degree22^{\prime}50\farcs70$.
To resolve the disk structure, we used the 12 m array with both extended (C43-6) and compact (C43-1) configurations; to recover the circumstellar disk/envelope, the 7 m Atacama Compact Array (ACA/Morita Array) was also employed. The combination of the 12 m and 7 m arrays resulted in baseline lengths ranging from 7 m to 2145.75 m.

\begin{figure}
    \centering
    \hspace{-0.9cm}
    \includegraphics[width=10cm]{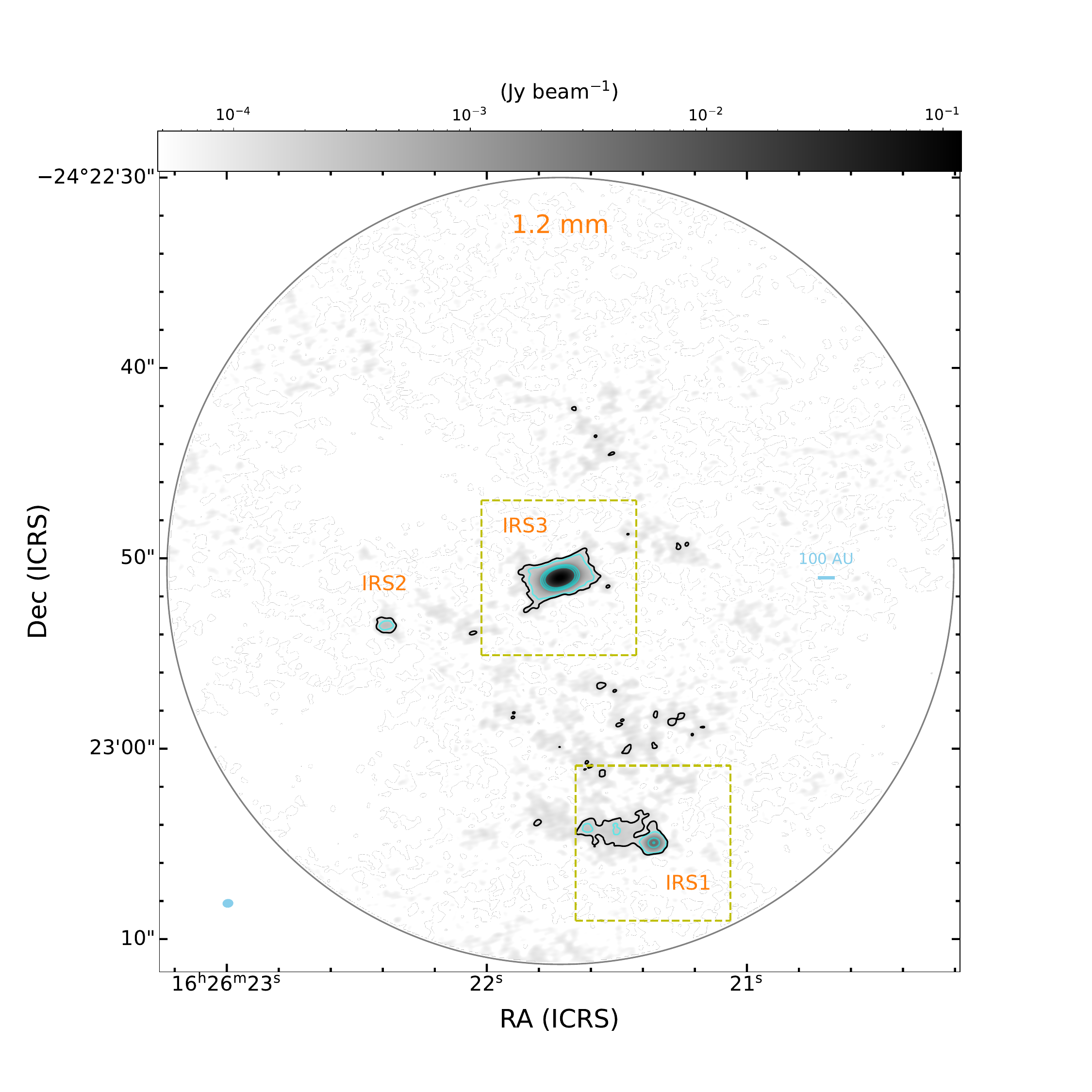}
    \caption{The continuum images obtained by combining the ALMA 12 m and 7 m arrays, using Briggs weighting of 2.0 to enhance the recovery of more extended emissions. Graymap shows the primary beam uncorrected continuum emission at 1.2\,mm (Setup2), while the 3$\sigma$ black contour shows the extend structure and cyan contours overlay emissions ranging of [5, 50, 100, 200]$\times\sigma$ ($\sigma$ = 47.6 $\mu$Jy beam$^{-1}$) for the compact region to keep clear in high gray-scales. 
    Blue ellipses in the lower-left corner of each panel represent the synthesized beam. The yellow dashed line boxes denote the zoom-in region shown at Figure~\ref{fig:cont}(a-f).}
    \label{fig:wfield}
\end{figure}

Our frequency setups cover the ranges 214.0–219.0 GHz and 229.0–234.0 GHz (Setup1), 242.5–247.5 GHz and 257.5–262.5 GHz (Setup 2), 92.0–96.0 GHz and 104.0–108.0 GHz (Setup 3). Within each spectral setup, one spectral window (spw) with a bandwidth of 1.875 GHz and a frequency resolution of 1.129 MHz was used for the continuum \citep[see ][]{2021FrASS...8..227C}. The bandpass and flux calibrators were J1256-0547, J1924-2914, J1517-2422, and J1427-4206; the complex gain calibrator was J1625-2527.
According to the ALMA Technical Handbook\footnote{\url{https://almascience.eso.org/documents-and-tools/cycle11/alma-technical-handbook}}, the 2$\sigma$ absolute flux calibration uncertainty is approximately 10\%.

Data reduction was performed using the Common Astronomy Software Applications (CASA) package version 5.6.1 \citep{2022PASP..134k4501C}. We applied a customized version of the ALMA calibration pipeline, supplemented by an additional in house routine to correct the $T_{\rm sys}$ and normalize the spectral line data. 
Self-calibration, based on line-free continuum emission, was employed for each configuration, aligning both amplitude and phase across multiple configurations. 

For Setup 1 and 2, we use 12m and 7m arrays, while for Setup 3 we use 12m only, achieving the maximum recoverable scales (MRS) as 34$^{\prime\prime}$, 28$^{\prime\prime}$, 14$^{\prime\prime}$ of Setup 1, Setup 2, Setup 3 respectively. 
For all configuration, the sources within the GGS30 field are strong enough to self-calibrate the phases on per-integration timescales in Setups 1 and 2, even in the most extended configuration used. For Setup3, the more extended configuration needed a phase solution interval of 15s because the sources are weaker, but also, Setup3 had better phase stability. Thus, the data presented do not suffer from decorrelation.

The self-calibration resulted in complex gain corrections, which were applied to all channels. Subsequently, the derived continuum model was subtracted to produce the line data, and images were generated using the {\tt tclean} task within CASA. We used the gridder = mosaic for combining 12m and ACA data (1.2\,mm and 1.3\,mm), and gridder = standard for 12m only (3.0\,mm). For continuum imaging, Briggs weighting with robust parameters of $-2.0$, 0.5, and 2.0 was tested (Table \ref{table:cont}).
Since the robust =2.0 recover the largest scale structure, in the following analyses are based on the images by using the robust=2.0, achieving the sensitivity of $\sigma$ = 47.6 $\mu$Jy beam$^{-1}$ at 1.2 mm, $\sigma$ = 49.4 $\mu$Jy beam$^{-1}$ at 1.3 mm, $\sigma$ = 46.0 $\mu$Jy beam$^{-1}$ at 3.0 mm before primary beam correction.
The primary beam of 1.2 mm 1.3 mm and 3.0 mm are 25$^{\prime\prime}$, 27$^{\prime\prime}$, 63$^{\prime\prime}$ respectively. The sensitivity at 50\% primary beam after the primary beam correction are list in Table \ref{table:cont}.

In general, the “JvM effect” should be considered in situations where there are only a small number of CLEAN beams across an image and the source flux may not be fully deconvolved \citep{1995AJ....110.2037J,2021ApJS..257....2C}. In such cases, some studies have discussed the use of a scaling factor to account for the residual flux \citep{2021ApJS..257....2C}. However, the exact value of this factor depends on the specific definition adopted for the integration of the dirty beam, including the radial extent over which the beam power is summed. In principle, when integrated over a sufficiently large area, the total power of the dirty beam approaches zero. 
For the FAUST project, we therefore focus on ensuring that the images are properly deconvolved, with emission cleaned down to the noise level within the CLEAN mask. In addition, the synthesized beam is dominated by the 12 m array data, such that any potential JvM effect is expected to be negligible.

\begin{figure*}
    \hspace{-0.9cm}
    \includegraphics[width=20cm]{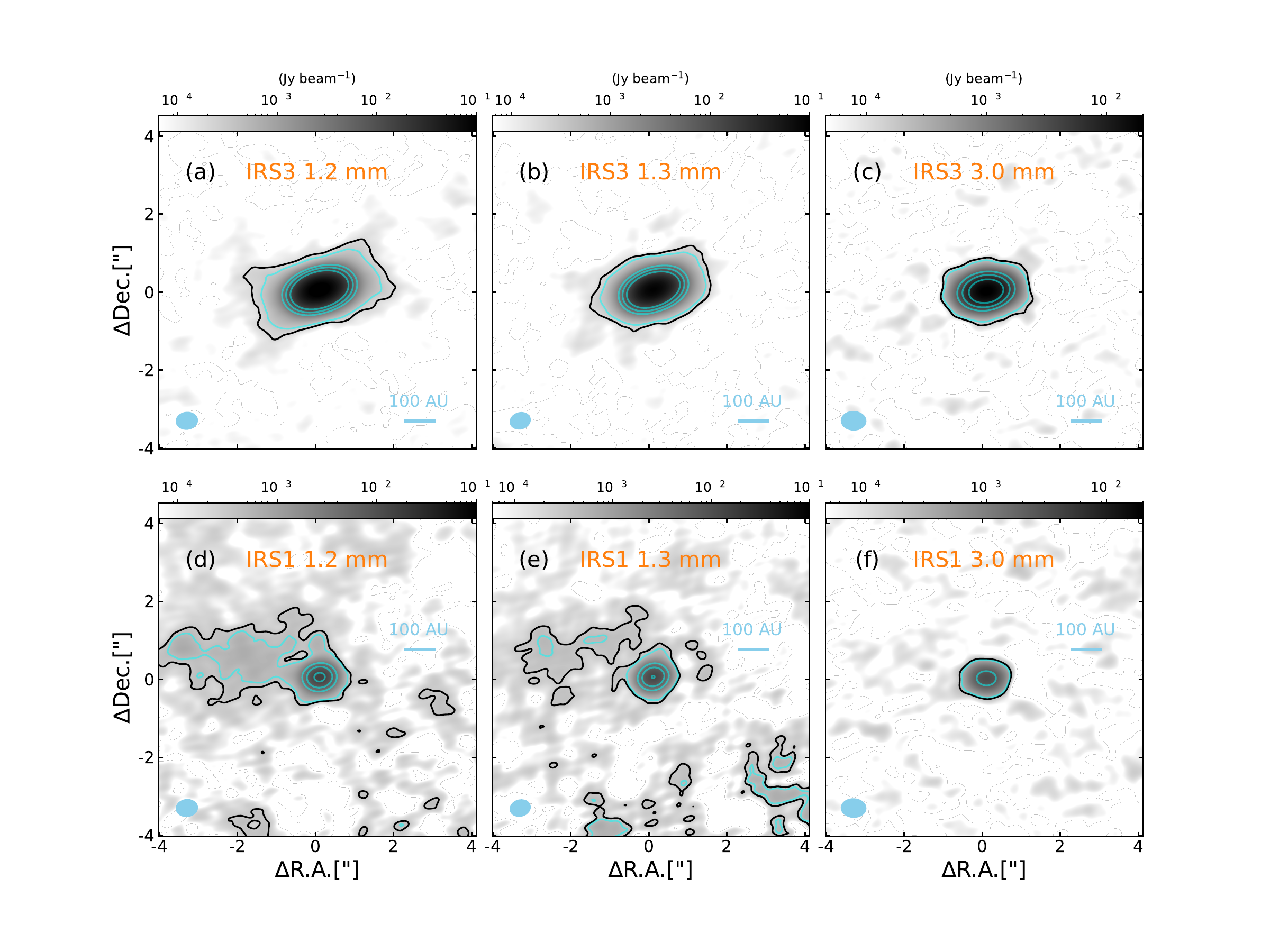}
    \caption{
    Graymap displays a zoomed-in view of the primary beam corrected continuum emission towards IRS3 and IRS1 at 1.2\,mm (Setup2), 1.3\,mm (Setup1) and 3.0\,mm (Setup3). The phase center (0,0) position is at coordinate according to Table \ref{table:cont}.
    (a)-(f) Using the same contours as Figure~\ref{fig:wfield} with the $\sigma$ (measured in the zoom-in image) of 64.4 $\mu$Jy beam$^{-1}$ at 1.2\,mm, 59.5 $\mu$Jy beam$^{-1}$ at 1.3\,mm and  45.6 $\mu$Jy beam$^{-1}$ at 3.0\,mm respectively. 
    Blue ellipses in the lower-left corner of each panel represent the synthesized beam.}
    \label{fig:cont}
\end{figure*}

\section{Results and analysis}\label{sec:result}

\subsection{Dust continuum images}\label{sec:continuum}

Figure \ref{fig:wfield} shows a primary-beam-uncorrected 1.2 mm image of GSS 30, within the 90\% primary beam response.
We detected the young stellar objects (YSOs) IRS1 and IRS3, with a signal-to-noise ratio larger than 100, in all three spectral setups; while IRS2 was detected at $>$7-$\sigma$ only in the Setup 2 ($\lambda\sim$1.2 mm).
Moreover, we found that IRS2 presents millimeter variability (more in Section \ref{sec:variability}).
We note that IRS2 was not detected in previous millimeter observations \citep{1997ApJ...475..713Z,2009A&A...507..861J,2018ApJ...869..158F}, except for the CO (2–1) emission reported by \citet{2024A&A...690A..46S}, which appears to be associated with the local cloud.

The peak positions of the continuum of these three sources at each wavelength were summarized in Table~\ref{table:cont}, and are consistent with previous findings from the \textit{Herschel} and VLA programs (\citealt{2013ApJ...770..123G,2013ApJ...775...63D}).

Figure \ref{fig:cont} shows the zoomed-in images of the resolved dust continuum.
In particular, in the 1.2 mm and 1.3 mm images, IRS1 appears as a spatially compact source, which is associated with an elongated structure extending towards the northeast; the elongated structure was not detected at $>$5-$\sigma$ in the 3.0 mm image (Figure \ref{fig:cont}d, e, f).
For IRS3 at the phase center, a $4\farcs6$ extended x-shape structure appears for emission above 3$\sigma$ at 1.2\,mm, shrinking into a box-shape at 1.3\,mm, and into an elliptical structure at 3.0\,mm.
The spatial extension of both IRS3 and IRS1 decreases with the wavelength increase. 

\subsection{Continuum spectra and spectral indices}\label{sec:spid}

To minimized systematic biases, the spectra and spectral index maps were evaluated from the continuum images that were created using a unified {\it uv} distance range of 12--380 $k\lambda$.
And we produced images for the spectral windows in the upper and lower sidebands separately, using the CASA-{\tt tclean} task with multi-frequency synthesis (mfs; \citealt{2011A&A...532A..71R}) nterms $=$ 1.
Before evaluating spectral indices, we smoothed all images to the identical synthesized beam of \beam=0\farcs6$\times$0\farcs6.
All sidebands images are showing Appendiex \ref{appendix:img}.

We measured flux densities ($F_{\nu}$) from the sideband images using two methods. 
First, we integrated flux densities from the innermost circular regions.
This strategy is suitable when the target sources are spatially resolved with subtle structures, which is our present case\footnote{At the liner resolution below 100 au, only eDisk and FAUST resolved this source. Therefore, in the following discussion, we compare our results with those reported in \citet{2024A&A...690A..46S}.} (see \citealt{2024A&A...690A..46S} for the ALMA images taken at further higher angular resolutions).
For IRS3, we used 4\farcs5 diameter to accommodate all the disk structures (Figure \ref{fig:cont}a, b, c); for IRS1, we used a 2$''$ diameter to focus on the emission of the spatially compact source instead of the eastern extension (Figure \ref{fig:cont}d, e, f).
Second, we measured flux densities by performing two-dimensional Gaussian fits (with only one component) using the CASA-{\tt imfit} task. 
Table \ref{tab:flux} summarizes the obtained flux densities. 
We take into account both thermal noise ($\sigma_{\rm th}$) and the absolute flux calibration errors ($\sigma_{\rm cal}$).
We nominally assumed $\sigma_{\rm cal}=0.1$ $F_{\nu}$.
The 1-$\sigma$ uncertainty of each flux density measurement was assumed to be $\sqrt{\sigma_{\rm th}^2 + \sigma_{\rm cal}^2}$.
The source, IRS2, presented  millimeter variability, which is described in Section \ref{sec:variability}.
The cases of IRS3 and IRS1 are discussed as follows.

\begin{table*}[htbp]
\centering
\caption{Flux density measurements.}
\label{tab:flux}
\begin{threeparttable} 
\begin{tabularx}{\textwidth}{lXXXX} 
\hline\hline
Frequency & \multicolumn{2}{c}{IRS3 Flux density (mJy)} & \multicolumn{2}{c}{IRS1 Flux density (mJy)} \\
(GHz) & Integrated from a region\tnote{a} & Gaussian fits\tnote{b} & Integrated from a region\tnote{c} & Gaussian fits\tnote{b} \\
& & & & \\
\hline
\multicolumn{5}{c}{Setup 3} \\
94.549  & 25.7 $\pm$ 0.28 & 24.9 $\pm$ 0.12 & 3.2 $\pm$ 0.14 & 3.1 $\pm$ 0.07 \\
106.150 & 33.1 $\pm$ 1.10  & 32.3 $\pm$ 0.27 & 3.5 $\pm$ 0.55 & 3.8 $\pm$ 0.29 \\

\multicolumn{5}{c}{Setup 1} \\
218.028 & 148 $\pm$ 0.79  & 143 $\pm$ 0.50 & 7.4 $\pm$ 0.40  & 6.9 $\pm$ 0.25 \\
232.910 & 170 $\pm$ 0.39  & 165 $\pm$ 0.50  & 8.1 $\pm$ 0.20  & 7.3 $\pm$ 0.26 \\

\multicolumn{5}{c}{Setup 2} \\
245.789 & 200 $\pm$ 0.35  & 193 $\pm$ 0.55 & 8.1 $\pm$ 0.18 & 7.2 $\pm$ 0.29 \\
259.949 & 222 $\pm$ 0.69  & 215 $\pm$ 0.67 & 8.3 $\pm$ 0.35 & 7.6 $\pm$ 0.27 \\
\hline
\end{tabularx}

\begin{tablenotes}
\raggedright
\item[a] Flux densities obtained by integrating pixel values over the inner 4\farcs5 (diameter) region.
\item[b] Flux densities obtained from 2D Gaussian fitting in the visibility domain.
\item[c] Flux densities obtained by integrating pixel values over the inner 2\farcs0 (diameter) region.
\item[d] The uncertainties listed include only thermal noise. The nominal absolute flux errors ($\sim 10\%$) should be considered in the scientific analysis.

\end{tablenotes}

\end{threeparttable}  
\end{table*}

\begin{figure*}
\sidecaption
    \includegraphics[width=12cm]{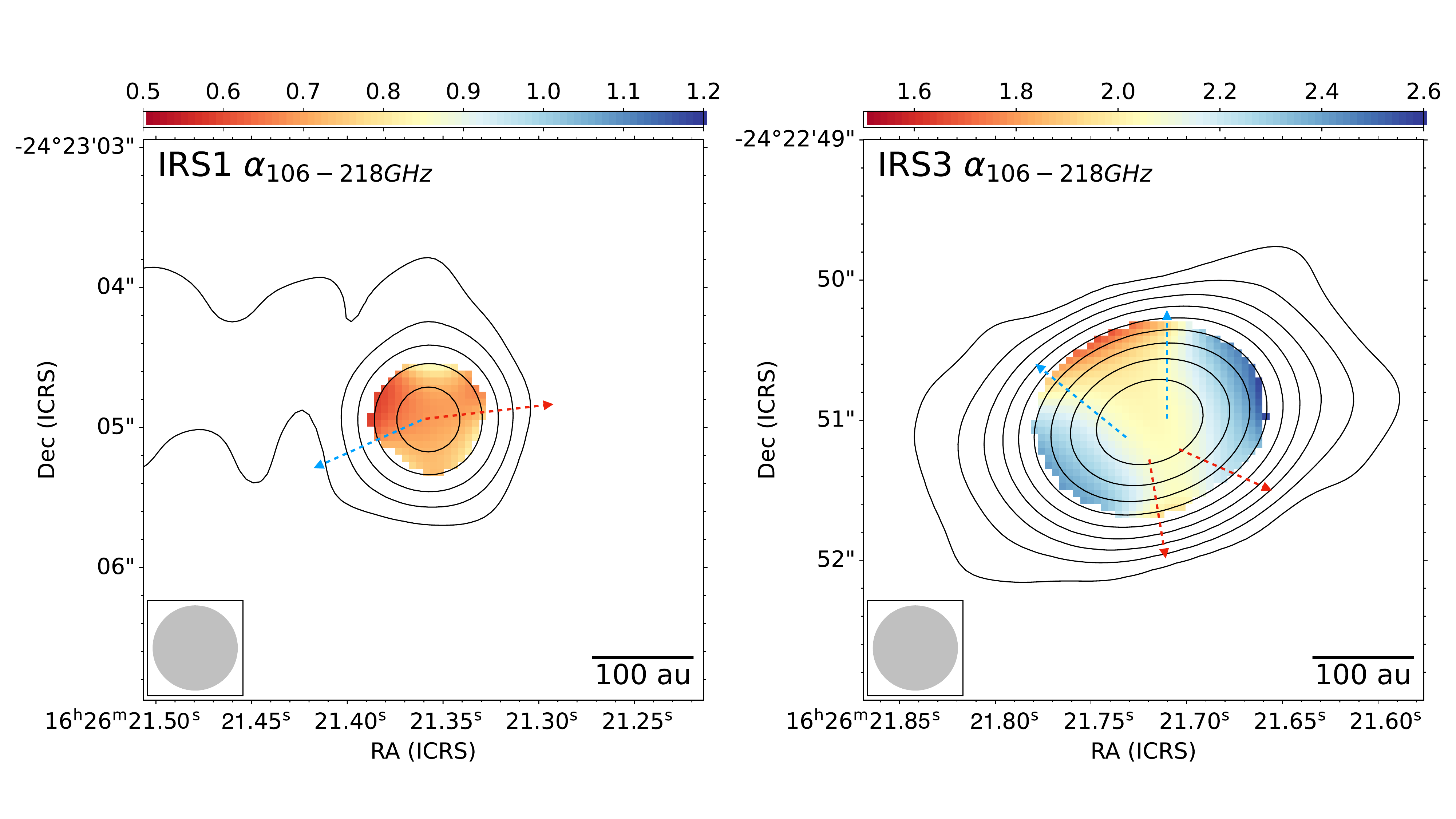}
    \caption{
    Spatial distribution of the 106--218 GHz spectral index ($\alpha_{\rm 106-218 GHz}$; color) and the 218 GHz intensity distribution (contours). All images were generated with circular beams with FWHM$=$ 0\farcs6. Contour levels are 276 $\mu$Jy\,beam$^{-1}$ (3-$\sigma$) $\times$[1, 2, 4, 8, 16, 32, 64, 128, 256]. The red and blue dashed arrows indicate the redshifted and blueshifted outflows traced by CCH (N = 3–2, J = 7/2–5/2, F = 4–3 and 3–2) emission (left) and SiO (5–4) emission (right), respectively, from Feng et al. (in prep.).
    }
    \label{fig:sedmap}
\end{figure*}

\paragraph{IRS3} We found that in Setup 3 (94--106 GHz), the flux densities measured using the two methods achieved good consistency; in Setups 1 and 2, the flux densities measured based on the first method are systematically higher than those measured based on the second method. 
Such discrepancy is likely due to the fact that the dust continuum images at $>$200 GHz frequencies trace spatially relatively extended features (Figure \ref{fig:cont}) which cannot be recovered by Gaussian fits with a single component. 
Therefore, in the following discussion for IRS3, we focus on the flux densities measured based on the first method. 
It is noteworthy that compared to the results of \citet{2024A&A...690A..46S}, our 225 GHz observation reveals a larger source size and our estimated flux density is higher than theirs by $\sim$23.6 mJy (18\%). 
This is likely because their observations at higher angular resolution have filtered out more extended emission.
The intra-band spectral index measured at 95--106 GHz $\alpha_{\rm 95-106~GHz}$ is $\sim$2.19$^{+0.37}_{-0.34}$.
The spectral index $\alpha_{\rm 106-218~GHz}$ is $\sim$2.08$^{+0.32}_{-0.28}$ in between 106 GHz and 218 GHz.
A linear regression for the $>$200 GHz data points (Table \ref{tab:flux}) yielded a $\alpha$ $\sim$2.36$\pm$0.14 spectral index, while a linear regression for all data yielded a $\alpha$ $\sim$2.12$\pm$0.14 spectral index.

\paragraph{IRS1} The flux density measurements obtained with the first method are generally higher than those obtained with the second method, likely due to the fact that the first method tends to include some dust emission in the eastern $\sim$500 au streamer-like structure. 
In addition, the first method yields a lower flux density at 106 GHz because the integrated region includes negative features around IRS1, which artificially reduce the total measured flux. 
In the following discussion for IRS1, we focus on the flux densities measured based on the second method. 
The intra-band spectral index measured at 95--106 GHz $\alpha_{\rm 95-106~GHz}$ is $\sim$1.6$^{+0.83}_{-0.87}$.
The spectral index $\alpha_{\rm 106-218~GHz}$ is $\sim$0.84$^{+0.16}_{-0.16}$ in between 106 GHz and 218 GHz.
A linear regression for the $>$200 GHz data points (Table \ref{tab:flux}) yielded a  0.48$\pm$0.16 spectral index, while a linear regression for all data yielded a 0.79$\pm$0.07 spectral index.

While the spectral indices are diagnostics in a spatially averaged sense, some more information can be derived from the spatially resolved spectral index maps.
We calculated the spectral index maps between 106 and 218 GHz at comparable high sensitivity and angular resolution.
Figure \ref{fig:sedmap} shows spatial distributions of 106--218 GHz spectral index ($\alpha_{\rm 106-218 GHz}$) for the two sources, IRS3 and IRS1. 
In IRS3, we found $\alpha_{\rm 106-218 GHz}\sim$2.0 around the peak of 218 GHz intensity; $\alpha_{\rm 106-218 GHz}$ is higher in the northwest and southeast ($\sim$2.5), and is lower in the northeast ($\sim$1.6).
In IRS1, the value of $\alpha_{\rm 106-218 GHz}$ is $\lesssim$0.8.

\subsection{Millimeter variability in IRS2}\label{sec:variability}

\begin{figure*}
    \centering
    \includegraphics[width=\textwidth]{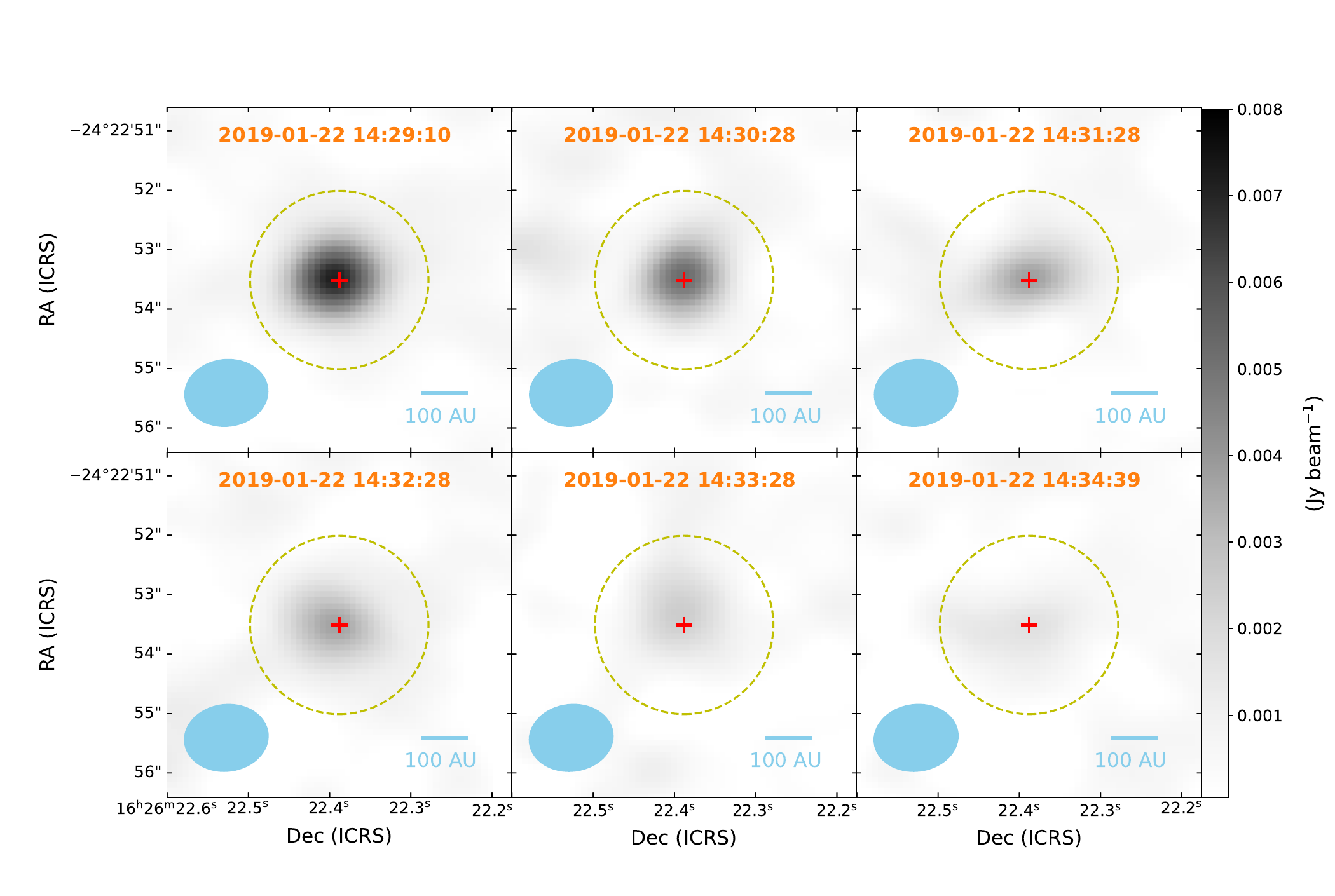}
    \caption{Images of IRS2 at 1.2\,mm with the time bin of 1 minute and the first and last frames having time bins of 98 s and 82 s, respectively. Red cross denote the IRS2 at center of $\alpha_{2000} = \rm 16h26m22s.39$, $\delta_{2000} = \rm -24\degree22^{\prime}53\farcs51$, and the yellow dashed circle has the radius of 1\farcs5. Sky-blue ellipses in the lower-left corner of each panel represent the synthesized beam ($1\farcs39 \times 1\farcs11$, PA = -83.8\degree).}
    \label{fig:time}
\end{figure*}

As mentioned in Section \ref{sec:continuum}, IRS2 was only detected with flux density $>3\sigma$ at 1.2\,mm. 
We noticed that this emission originated during the 420 seconds observations on January 22, 2019, while it was not significantly detected on other dates (see Table \ref{table:a1}).
Using a time bin of 1 minute (with the first and last frames having time bins of 98 s and 82 s, respectively), we obtained the images shown in Figure \ref{fig:time}, where the peak intensity gradually decreases from 7.17 $\pm$ 0.26 mJy beam$^{-1}$ to 1.69 $\pm$ 0.30 mJy beam$^{-1}$.

\begin{figure}
    \hspace{-0.3cm}
    \includegraphics[width=9cm]{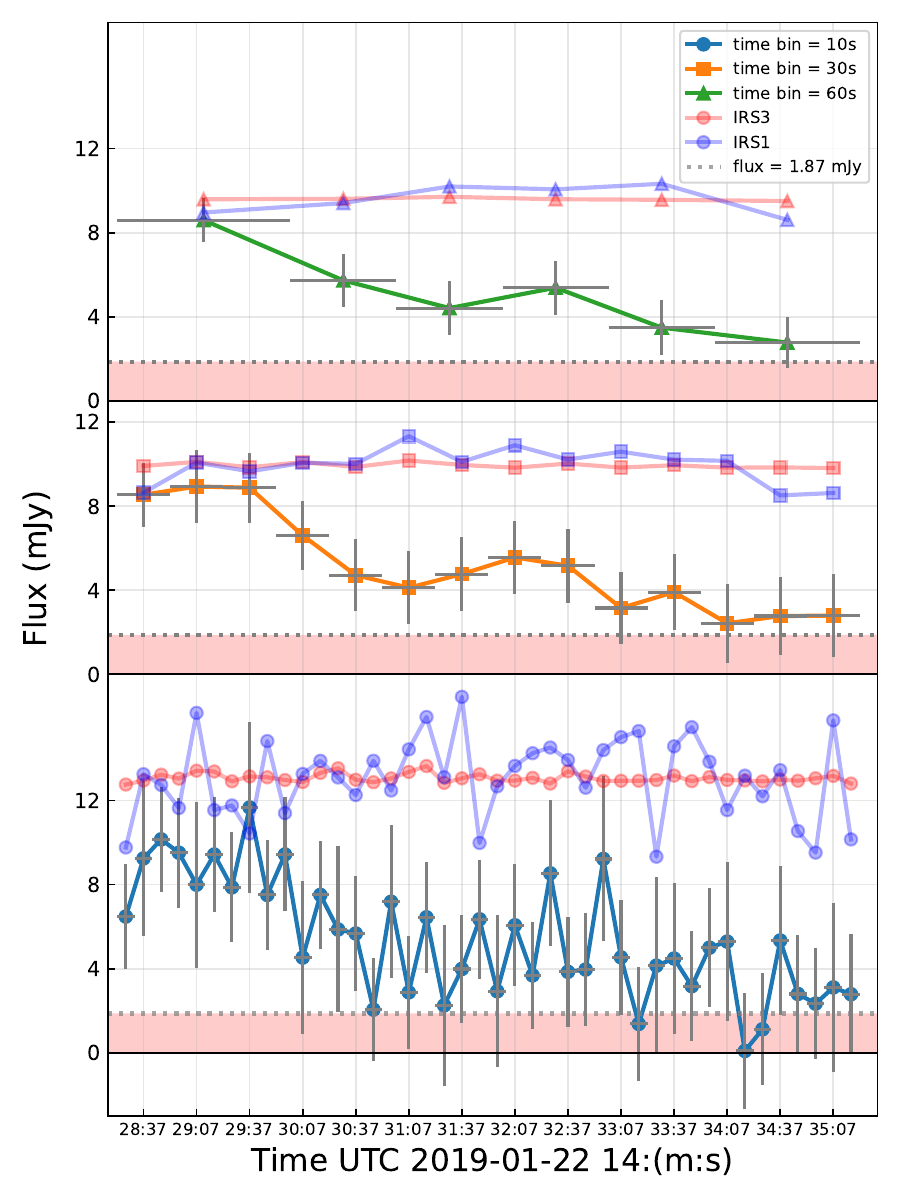} \\
    \caption{1.2\,mm light curve of IRS2, integrated from the 1\farcs5 circular region (yellow dashed circle in Figure~\ref{fig:time}) with the time bin of 10s, 30s and 60s respectively. The error bars show the 1$\sigma$ range of each images for y direction (flux) and integrated time range of each data points for the x direction (time). The gray dashed line and red shaded region represent the flux upper limit of combination from the observations excluding the time interval shown here. The red and blue lines show the light curves of IRS3 and IRS1, respectively, normalized to their mean fluxes.}
    \label{fig:light_curve}
\end{figure}

We further tested the light curve analysis with time bins of 30 s and 10 s by integrating the flux within a circular region with a radius of 1\farcs5 centered at intensity peak according to Table \ref{table:cont} (indicated by the red cross and yellow dashed circle in Figure \ref{fig:time}). The resulting light curves are shown in Figure \ref{fig:light_curve}. An upper limit of the flux from IRS2 was derived by imaging the data integrated from the rest of the observations at the same wavelength and integrating toward this region (indicated by the red shaded area in Figure \ref{fig:light_curve}).

The 1-minute binned light curve reveals that the integrated flux decreases from 8.54 $\pm$ 1.05 mJy to 2.73 $\pm$ 1.22 mJy, which may be related to a flare event. For the light curves with the shorter than 10s bin, the uncertainty smeared the credibility of the flux fluctuation; for light curves with longer than 1-minute bins, the detail of variability is lost. The $\sim$420 s exposure time is insufficient to capture a full cycle. In contrast, the 30 s binned light curve offers a more favorable balance between time resolution and signal-to-noise ratio. It suggests that the overall decay timescale likely exceeds 420 s and may include two decreasing sine-shaped waves with a period of $\sim$240 s.

The 1-$\sigma$ uncertainty of the intra-band spectral index was 6 due to the limited frequency coverage provided by one spectral setup (Section \ref{sec:observations}).

\subsection{Modeling spectral energy distributions}\label{sec:model}

For a tentative yet quantitative interpretation of the observed spectra (Figure \ref{fig:sed}), we produced simplified SED models that include dust emission components and free-free emission components.
Since our modeling involves the radiative transfer process within an optically thick environment, it is performed on the spatially-integrated spectra of the entire source (Table \ref{tab:flux}) rather than the resolved spectral index map (Figure \ref{fig:sedmap}).
We evaluated spectral profiles of free-free emission components based on the formulation presented in \citet{2003ApJ...599.1196K} and \citet{1967ApJ...147..471M}.
When evaluating the dust spectral profile, we employed the analytic solution of the Equations of radiative transfer presented as Equation (10)--(20) in \citet{2018ApJ...869L..45B} that took dust emission, absorption, and scattering into consideration.

We assumed that there is one dominant free-free emission source which has uniform electron density and temperature. 
In this case, the spectrum of free-free emission can be parameterized with the following three free parameters, the electron temperature ($T_{\rm e}$), emission measure ($EM$), and the solid angle of the free-free emission source ($\Omega_{\rm ff}$). 
Similarly, we assumed that the observed dust emission in each source is dominated by one dust emission source which has uniform dust temperature ($T_{\rm dust}$), dust volume density ($\rho_{\rm dust}$), maximum grain size ($a_{\rm max}$).
In this case, the spectrum of dust thermal emission can be parameterized with the following four free parameters, $T_{\rm dust}$, $a_{\rm max}$, dust column density ($\Sigma_{\rm dust}$), and the solid angle of the dust emission source ($\Omega_{\rm dust}$).

We adopted the default DSHARP dust opacity (\citealt{2018ApJ...869L..45B}).
When evaluating the size-averaged dust opacity, we assumed the grain size distribution $n(a)\propto a^{-3.5}$ in between the minimum and maximum grain sizes ($a_{\rm min}$, $a_{\rm max}$).
The size-averaged opacity depends weakly on the value of $a_{\rm min}$, which was assumed to be $10^{-4}$ mm. 
We optimized the value of $a_{\rm max}$ according to the observational data.
We refer to the Appendix C of \citet{2024ApJS..273...29C} for a comprehensive discussion of the effects of adopting other commonly used opacity tables.

Approximating the free-free subtracted flux densities observed at $>$12 GHz with the spectral profile of an isothermal dust emission slab, we can determine the values of $a_{\rm max}$ and dust column density ($\Sigma_{\rm dust}$) in the slab by performing a Markov chain Monte Carlo (MCMC) fit, using the {\it emcee} package. 
The best-fit parameters are summarized in Table \ref{table:bf}.

\begin{table}[!htbp]
\caption{Best fit parameters of SED model.}
\centering
\label{table:bf}
\begin{threeparttable} 
\begin{tabularx}{0.5\textwidth}{XXX} 
\hline\hline
Source &  IRS3 & IRS1 \\
\hline
$T_{\rm e}$\tnote{a} (K) & 8000 & 8000 \\
$EM$\tnote{b} ($\rm cm^{-6}$ pc) & $10^8$ & $10^{10}$ \\
$\Omega_{\rm ff}$\tnote{b} (sr) & 5$\times10^{-14}$ & 5$\times10^{-15}$ \\
$T_{\rm dust}$\tnote{a} (K) & 49 & 18 \\
$\alpha_{\rm max}$\tnote{b} ($\mu$m) & 30$_{-18}^{+12}$ & $\sim$150\tnote{c} \\
$\Sigma_{\rm dust}$\tnote{b} (g $\rm cm^{-2}$) & 24.5$_{-5.2}^{+8.2}$ & >5\tnote{c} \\
$\Omega_{\rm dust}$\tnote{a} (sr) & 2.4$\times10^{-12}$ & 2.5$\times10^{-13}$ \\
\hline
\end{tabularx}

\begin{tablenotes}
\raggedright
\item[a] Fixed parameters.
\item[b] Fitted parameters.
\item[c] More detail in discussion in Section\,\ref{sub:irs1}.

\end{tablenotes}

\end{threeparttable}  
\end{table}

\section{Discussion}\label{sec:discussion}

In this section, we discuss the results and provide an interpretation for the resolved morphology, SEDs, and spectral index maps in IRS3 and IRS1, and the millimeter variability observed in IRS2.

\subsection{IRS3}\label{sub:irs3}

As compared with \citet{2024A&A...690A..46S}, the ALMA data we used for the present work has larger synthesized beams and larger MRS. As a result, we recovered some emission structures that have lower surface brightness, which may include the free-free emission source along the disk minor axis, and the optically thinner outer disk along the disk major axis. 
They result in the $<$2 and $>$2 spectral indices resolved in the left panel of Figure \ref{fig:sedmap}. 
Moreover, the $\alpha_{\rm 106-218 GHz}$ is below 2.0 around the regions that spatially coincide with the outflow cavity walls traced by the CCH (N = 3–2, J = 7/2–5/2, F = 4–3 and 3–2) emission (indicated by dashed arrow in Figure~\ref{fig:sedmap}, from Feng et al. in prep.) and $^{12}$CO (2–1) emission (e.g., Figure 6 in \citealt{2024A&A...690A..46S}); $\alpha_{\rm 106-218 GHz}$ is higher in the ambient region.
As a result, the spatially averaged spectral indices at 95--218 GHz is consistent with 2, while it is slightly higher at higher frequencies (Section \ref{sec:spid}).

\begin{figure}[h]
    \hspace{-0.3cm}
    \begin{tabular}{c}
         \includegraphics[width=9cm]{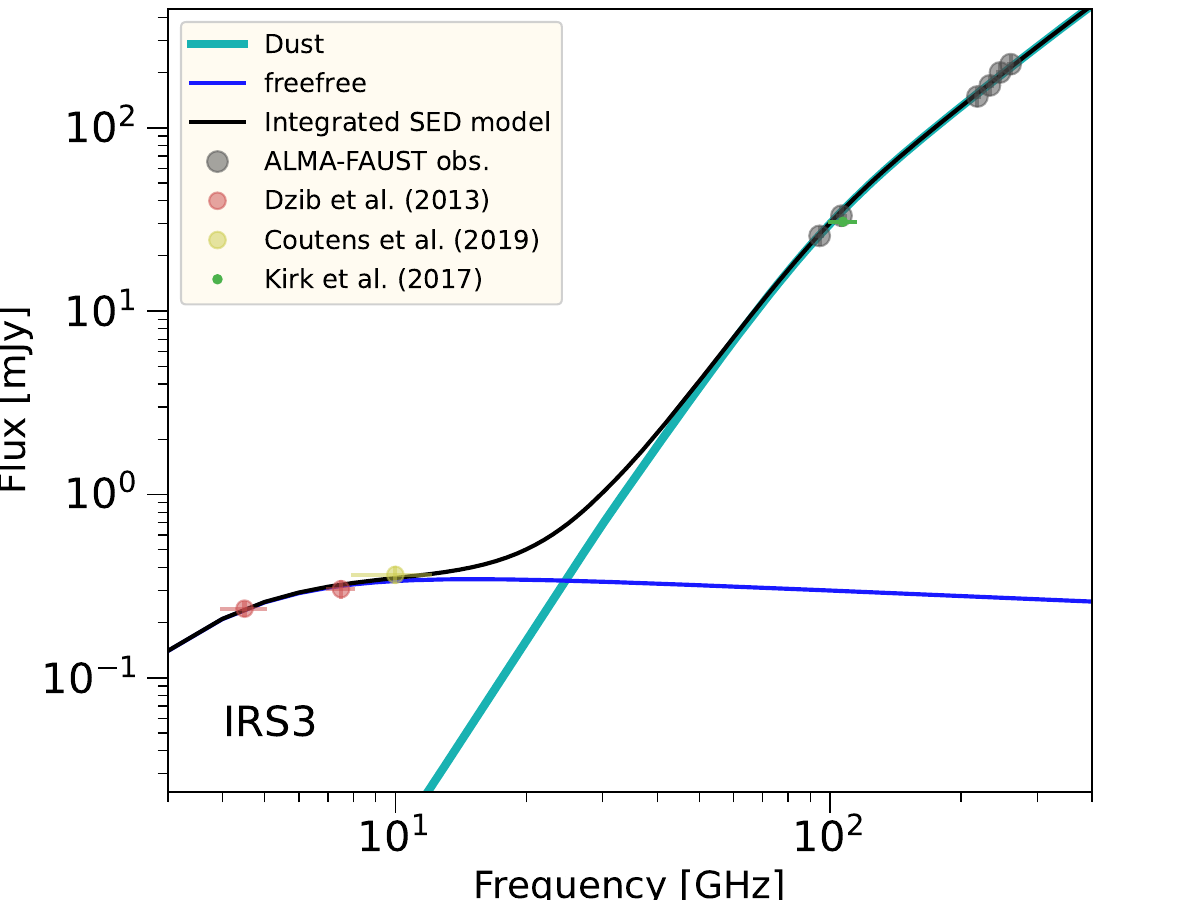} \\
         \includegraphics[width=9cm]{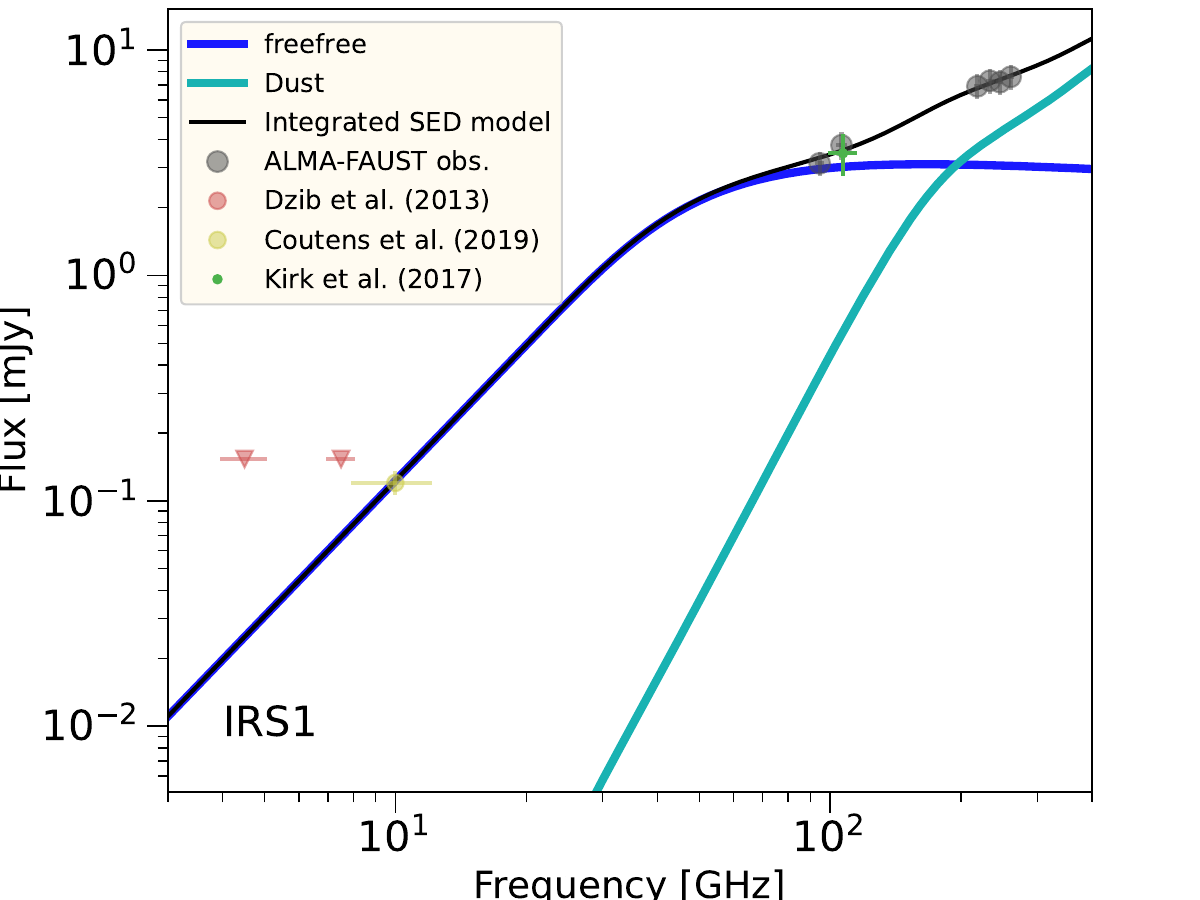} \\
    \end{tabular}
    \caption{
    The observed spectral energy distributions (SED; symbols), overplotted with the predictions from our simplified SED model (solid lines. We quote the observations reported in \citet{2013ApJ...775...63D,2017ApJ...838..114K,2019A&A...631A..58C}. The data points show in full circle, and the inverted triangle denote the upper limit.
    }
    \label{fig:sed}
\end{figure}

Since the previous JVLA X (4.5 GHz, 7.5 GHz) and C (10.0 GHz) band observations (Figure \ref{fig:sed}; \citealt{2013ApJ...775...63D,2019A&A...631A..58C}) have constrained the free-free emission (\citealt{1986ApJ...304..713R}) to be more than 2 orders of magnitude fainter than dust emission at $>$100 GHz (more below), the lower than 2 spectral indices can be associated with the following three mechanisms: (1) the observations of dust thermal emission were not entirely in the Rayleigh-Jeans limit (\citealt{1983QJRAS..24..267H}), (2) the maximum dust grain size $a_{\rm max}$ is $\sim$0.1 mm and the dust spectral profile is anomalously reddened due to dust self-scattering (\citealt{2019ApJ...877L..22L}), and (3) dust emission from the inner, higher temperature region is obscured by some lower temperature dust in the foreground (e.g., the outer disk or inner envelope; \citealt{2017ApJ...840...72L,2018ApJ...868...39G,2019ApJ...884...97L}).

In the following, we discuss these three possibilities based on a simple SED model that include one dust emission source and one free-free emission source by adopted the default DSHARP dust opacity (\citealt{2018ApJ...869L..45B}).

The flux densities observed at $<$12 GHz are consistent with free-free emission with an electron temperature $T_{e}=$ 8000 K, an emission measure $EM=$ 10$^{8}$ cm$^{-6}$\,pc, and a solid angle $\Omega=$ 5$\cdot$10$^{-14}$ sr ($\sim$2.1$\times$10$^{-3}$ arcsec$^{2}$.
This solid angle corresponds to a circular area with $\sim$7 au diameter), although these parameters remain degenerate and might have been biased due to the potential radio variability (e.g., \citealt{2013ApJ...775...63D,2014ApJ...780..155L}).  
The remaining flux density at $>$12 GHz frequencies is likely dominated by dust thermal emission.

Following \citet{2024A&A...690A..46S}, based on the luminosity of GSS30, we assumed a passively heated averaged dust temperature of $T_{\rm dust}=$ 49 K.
In this case, dust emission at 106--218 GHz is in the Rayleigh-Jeans limit, which disfavored interpreting $\alpha_{\rm 106-218 GHz}<$2 with the aforementioned mechanism (1).

Previous high angular resolution images have constrained the disk size to 76$\times$23 au, corresponding to the solid angle ($\Omega$) of the dust emission source to be $\sim$2.4$\times$10$^{-12}$ sr (evaluated based on the full width at half maxima in the major and minor axes of the GSS 30 IRS3 disk reported in \citealt{2024A&A...690A..46S}).
We found that with the $T_{\rm dust}=$49 K assumption and this solid angle, the observed flux densities at 95--260 GHz (Figure \ref{fig:sed}) that cannot be reproduced if $a_{\rm max}$ is comparable or greater than 0.1 mm, due to strong attenuation by dust scattering.
Therefore, we also disfavored interpreting $\alpha_{\rm 106-218 GHz}<$2 with aforementioned mechanism (2).

At 218--260 GHz, the spectral index of our SED model is lower than what was measured (Figure \ref{fig:sed}). 
This is likely because the column density assumption of our model is over-simplified.
This can be amended by assuming a radially decreasing density profile.
In this case, the optically thinner emission in the outer regions can yield high spatially-integrated spectral indices at high frequencies (c.f. \citealt{2019ApJ...884...97L}) and the $\sim$2.5 values of $\alpha_{\rm 106-218 GHz}$ resolved in the northwest and southeast of the spectral index map (Figure \ref{fig:sedmap}).

Moreover, the temperature assumption of our SED model is also over-simplified.
Given the high averaged optical depth in this model, assuming a radially decreasing temperature profile will allow interpreting the $<$2 values of $\alpha_{\rm 106-218 GHz}$ resolved in the spectral index map (Figure \ref{fig:sedmap}) with the aforementioned mechanism (3).
For example, the surfaces of the outflow cavity or the embedded, inner (pseudo-)disk may have higher dust temperature than the outer disk and/or the ambient gas envelope. 
This implies that the spatially resolved variation of $\alpha$ is a natural manifestation of the high optical depth and temperature gradients. 
While our SED model adopts a simplified isothermal assumption, the observed local reduction in $\alpha$ along the minor axis primarily reflects self-obscuration, which is consistent with the high optical thickness identified by our modeling.
We note that the previous higher angular resolution 225 GHz continuum image has spatially resolved weak asymmetry along the disk minor axis (\citealt{2024A&A...690A..46S}), which is expected for a self-obscured dusty disk with a radially decreasing temperature profile (\citealt{2018ApJ...868...39G}).

The dust mass in our simple SED model is $M^{\rm slab}_{\rm dust}\sim$1510$^{+636}_{-243}$ $M_{\oplus}$, which is considerably higher than the estimates of \citet{2024A&A...690A..46S} (24--70 $M_{\oplus}$) that were made based on the optically thin assumption and assuming the dust absorption opacity $\kappa_{\rm 225 GHz}=$ 2.3 cm$^{2}$\,g$^{-1}$. 
A part of the difference is due to the relatively low $\sim$250 GHz dust absorption opacity ($\sim$0.5 cm$^{2}$\,g$^{-1}$) in the DSHARP table (\citealt{2018ApJ...869L..45B}).
With the the best-fits are $a_{\rm max}=$ 30$^{+12}_{-18}$ $\mu$m and $\Sigma_{\rm dust}=$ 24.5$^{+8.2}_{-5.2}$ g\,cm$^{-2}$, the dust optical depth at $\sim$1 mm wavelength may be as high as 25, implies that the dust luminosity at $>$200 GHz is dominated by very optically thick dust structures.
Such $a_{\rm max}$ and $\Sigma_{\rm dust}$ values are reasonable as compared to the analyses of the recent multi-frequency dust polarization observations towards Class 0/I objects (\citealt{2020ApJ...889..172K,2021ApJ...914...25L,2024A&A...682A..56Z}).

As mentioned above, the spectral index distribution in Figure \ref{fig:sedmap} can be explained by a hot dust emission source that dominates the flux densities at $\lesssim$100 GHz, but which is obscured at higher frequencies by a cooler dust emission source that is optically thick at 218--260 GHz. 
In this case, the $\Sigma_{\rm dust}$ of the cooler, obscuring dust emission source can be considerably lower than 23 g\,cm$^{-2}$ since it only needs to be optically thick at $>$200 GHz.
The obscured hot dust emission source does not need to harbor a lot of dust mass since having a high $T_{\rm dust}$ makes it an efficient emitter.
Therefore, $M^{\rm slab}_{\rm dust}$ should be regarded as an upper limit of dust mass in the IRS3 disk. 
Nevertheless, we found that assuming $T_{\rm dust}=$ 49 K, making the cooler, obscuring dust emission source optically thick enough to match the observed flux densities at 218--260 GHz, requires  $\Sigma_{\rm dust}>$10 g\,cm$^{-2}$.
In this case, $M^{\rm slab}_{\rm dust}$ may be as low as $\sim$650 $M_{\oplus}$.

The dust mass in IRS3 is dominated by the region resolved in \citet{2024A&A...690A..46S}, which is also the same region assumed in our model. While our maps reveal spatial variations in the spectral index, our mass estimate is derived from the integrated SED, which provides a global constraint on the total emission budget. Our present dust mass estimate for IRS3 has already taken free-free emission into consideration. Since the 225 GHz flux density we detected is only 18\% higher than what was reported in \citet{2024A&A...690A..46S}, any local excess from a potential ionized jet or errors from neglecting the optically thinner outer disk are effectively capped by this flux difference. Thus, the impact of these spatial inhomogeneities on our total dust mass estimate is expected to be less than 18\%.

Based on our dust mass estimate, we make a tentative assessment of the gravitational stability of the IRS3 disk using the Toomre Q parameter \citep{1960AnAp...23..979S,1964ApJ...139.1217T}. Assuming a standard gas-to-dust ratio of 100 and that the gas temperature is equal to the dust temperature ($\sim$49 K), we obtain Q $\simeq$ 0.16.
Under these assumptions, the inferred disk mass could be comparable to or even exceed the stellar mass (0.35 $M_{\odot}$; \citealt{2024A&A...690A..46S}),which may suggest that the disk is gravitationally unstable. 
However, the interpretation of the Toomre Q parameter is most straightforward for approximately isothermal disks, whereas Class 0/I disks are unlikely to be strictly isothermal.
In addition, such a mass configuration may be stable for transients, as suggested by numerical simulations (e.g., \citealt{2010ApJ...708.1585K, 2013MNRAS.431.1719M}).
It is also important to note that this apparent instability could be sensitive to the adopted assumptions.
In an extreme case, this mass configuration can remain stable with a Toomre Q parameter of 1–3, provided that the gas temperature exceeds 200 K and the gas-to-dust ratio is below 30 which is consistent with typical disk-to-stellar mass ratios in recent simulations (e.g., \citealt{2024A&A...686A.253M}).
Given the current uncertainties in profiles, the stability of the IRS3 disk remains poorly constrained.

\subsection{IRS1}\label{sub:irs1}

The emission mechanisms in the compact source in IRS1 is not well constrained by the present observations (Figure \ref{fig:sed}, bottom).
Using SED modeling, we tried to address what could be the dust mass budget in this system if the emission at 95 GHz and at lower frequencies is dominated by free-free emission.
In this case, the present 95 GHz data and the previous X band observations (\citealt{2019A&A...631A..58C}) constrained the turnover frequency of the free-free emission and thereby the emission measure, while the electron temperature and the solid angle of this free-free emission source are degenerate.
Assuming $T_{e}=$ 8000 K, we found that the observations are consistent with $EM=$ 10$^{10}$ cm$^{-6}$\,pc and $\Omega=$ 5$\cdot$10$^{-15}$ sr.
To explain the remaining flux densities at $>$95 GHz with dust emission, it requires the dust emission source to be relatively cool (i.e., not in the Rayleigh-Jeans limit), optically thick, and the spectra at $\sim$200 GHz need to be anomalously reddened by dust self-scattering (\citealt{2019ApJ...877L..22L}).
Otherwise, it is not possible to reproduce the observed low spectral indices. 
We find that it is consistent with $T_{\rm dust}=$ 18 K, $a_{\rm max}=$ 0.15 mm, and $\Omega=$ 2.5$\times$10$^{-13}$ sr ($\sim$10$^{-2}$ arcsec$^{2}$), with $\Sigma_{\rm dust}>$5 g\,cm$^{-2}$.
The lower limit of dust mass is 32 $M_{\oplus}$.
The $T_{\rm dust}$ value provided here is a weak upper limit, which may be realistic since it can be achieved due to the interstellar radiation field. 
When assuming lower $T_{\rm dust}$ values, the $a_{\rm max}$ values can be larger or smaller than 0.15 mm while the value of $\Omega$ will be considerably larger than 2.5$\times$10$^{-13}$ sr, leading to a larger dust mass estimate. 
It is possible to model the $>$95 GHz flux density in IRS1 with another free-free emission component although its turnover frequency and $EM$ may be unexpectedly high.

In the 1.2 and 1.3 mm images (Figure \ref{fig:cont}), we note that IRS1 exhibits an extended structure toward the northeast, spanning 370–500 au, which is also present in the ALMA dust polarization survey by \citet{2019ApJS..245....2S}. The same feature was also detected in the scattered light observations with the Very Large Telescope (VLT)/NACO, which further revealed IRS1 as a protobinary system (e.g., Figure 1 in \citealt{2007A&A...475..277C}). Moreover, in the SiO (5–4) emission (Feng et al., in prep.), the angle between the redshifted and blueshifted lobes is about 160$\degree$ (Figure \ref{fig:sedmap}, right), which is nearly perpendicular to both the CO (2–1) emission reported by \citet{2018ApJ...869..158F} and the dust polarization direction \citep{2019ApJS..245....2S}. Taken together, the aforementioned observational results suggest that the extended continuum emission may be part of an outflow–disk complex or a trailing structure associated with a circumbinary disk around the two YSOs.

Given that the extended structure is morphologically similar to the streamers or flyby-induced features reported in recent millimeter line and continuum observations (e.g., \citealt{2016SciA....2E0875L,2018A&A...612A..54L,2022NatAs...6..331D,2022NatAs...6..837L,2023ApJ...951...11Y,2024A&A...682A..61C}), we cannot exclude the possibility that it represents an accretion flow or a tidally induced arm in the absence of kinematic information from line data.
We will leave this possibility to be tested in the future observations.

\subsection{IRS2}\label{sub:irs2}

IRS2 is classified as a weak-lined T~Tauri star \citep{1959SvA.....3..434D}. 
Although our observations might not have a long enough duration to capture its full cycle of millimeter variability, significant variation of millimeter flux density was indeed detected over the observations with a 420s duration. 
In addition, multiple snapshot observations between November~2018 and September~2019---both before and after January~22---failed to detect any >3$\sigma$ emission from IRS2 (1.2\,mm, 2018-11-17, <5.44 mJy beam$^{-1}$ for 7\farcs4 $\times$ 3\farcs9; 2018-11-29, <0.28 mJy beam$^{-1}$ for 0\farcs6 $\times$ 0\farcs3; 2019-04-19, <0.18 mJy beam$^{-1}$ for 0\farcs5 $\times$ 0\farcs5; 1.3\,mm and 3.0\,mm see Table \ref{table:cont}). 
Moreover, previous X-ray monitoring observations of \textit{Chandra} showed that IRS2 varied on timescales of a few thousand seconds \citep{2003PASJ...55..653I}.
The observed 1.2\,mm flux variability amplitude of IRS2, measured as the difference between the highest and lowest fluxes normalized by the peak flux, is $68.0_{-20.8}^{+16.2}$\%. 
This is consistent with the variability reported in previous VLA observations \citep{2013ApJ...775...63D}, which showed flux variations of 68.6\% at 4.5\,GHz and 85.1\% at 7.5\,GHz.
These multi-wavelength variabilities can be interpreted as a result of magnetic flaring activity at the surface of the protostar \citep{2002ARA&A..40..217G}. In this scenario, the mm emission must be non-thermal and produced by relativistic electrons accelerated during magnetic reconnection events. To produce detectable mm emission, the electrons must attain substantial Lorentz factors \citep{2006A&A...453..959M}, implying a highly energetic acceleration process.

\section{Conclusions}\label{sec:conclusions}

The present study provides a detailed continuum analysis of three YSOs within the GSS 30 complex, as part of the ALMA FAUST large program. 
Observations at 1.2 mm, 1.3 mm, 3.0 mm enabled, for the first time, detailed spectral index mapping and SED analysis of these protostellar envelope/disk systems.

IRS3 – The spectral index $\alpha_{\rm 106-218~GHz}$ increases from the center (2.0) to the disk edge along the radii (2.5) and decreases along the outflow direction perpendicular to the disk (1.6-1.8). In particular, an asymmetric distribution is noted in the spectral index map, with $\alpha$ as small as 1.6 to the northeastern blue shifted lobe. Colder dust in the outer envelope obscuring warmer dust in the inner region may be the reason. From SED fitting, the maximum dust grain size is tens of micros and the dust mass is 650-1510 M$_\oplus$.

IRS1 – Besides the central compact protostar, we noted an extended structure to the northeast of IRS1 possibly related to part of an outflow–disk complex or a trailing structure associated with a circumbinary disk. Contamination from free-free emission likely suppresses the spectral index towards the compact protostar, yielding an upper limit of 0.8.

IRS2 – A brightness variation was observed within a total exposure time of 420s. By comparing observations from different epochs, we infer that the source likely underwent a flare lasting on the order of tens of minutes. This timescale suggests a connection with magnetic activity in the central protostar.

\begin{acknowledgements}
Q.Y. and S.F. acknowledges support from the National Key R\&D program of China grant (2025YFE0108200), National Science Foundation of China (12373023,   12133008). 
H.B.L. is supported by the National Science and Technology Council (NSTC) of Taiwan (113-2112-M-110-022-MY3). 
S.V and M.B. acknowledge the support from the European Research Council (ERC) Advanced grant MOPPEX 833460. LL acknowledges the support of DGAPA-UNAM project PAPIIT IN108224 and CONAHCyT project CF-263356. 
E. B. acknowledges contribution of the Next Generation EU funds within the National Recovery and Resilience Plan (PNRR), Mission 4 - Education and Research, Component 2 - From Research to Business (M4C2), Investment Line 3.1 - Strengthening and creation of Research Infrastructures, Project IR0000034 – “STILES - Strengthening the Italian Leadership in ELT and SKA”. 
E. B. also acknowledges support from the Italian Ministry for Universities and Research under the Italian Science Fund (FIS 2 Call – Ministerial Decree No. 1236 of 1 August 2023), grant FIS-2023-00170.
G.S., Cl.Co. and L.P. acknowledge financial support under the National Recovery and Resilience Plan (NRRP), Mission 4, Component 2, Investment 1.1, Call for tender No. 104 published on 2.2.2022 by the Italian Ministry of University and Research (MUR), funded by the European Union – NextGenerationEU-Project Title 2022JC2Y93 Chemical Origins: linking the fossil composition of the Solar System with the chemistry of protoplanetary disks – CUP J53D23001600006 – Grant Assignment Decree No. 962 adopted on 30.06.2023 by the Italian Ministry of Ministry of University and Research (MUR). 
G.S., Cl.Co. and L.P.  also acknowledge the project ASI-Astrobiologia 2023 MIGLIORA (“Modeling Chemical Complexity”, F83C23000800005), the INAF-GO 2024 fundings ICES, the INAF-GO 2023 fundings PROTOSKA (“Exploiting ALMA data to study planet forming disks: preparing the advent of SKA”, C13C23000770005), the INAF Mini-Grant 2022 “Chemical Origins” (PI: L. Podio) and the INAF Minigrant 2023 TRIESTE (“TRacing the chemIcal hEritage of our originS: from proTostars to planEts”; PI: G. Sabatini). 
SBC was supported by the NASA Planetary Science Division Internal Scientist Funding Program through the Fundamental Laboratory Research work package (FLaRe). 
I.J-.S acknowledges funding from grant PID2022-136814NB-I00 funded by the Spanish Ministry of Science, Innovation and Universities/State Agency of Research MICIU/AEI/ 10.13039/501100011033 and by “ERDF/EU”. FMe has received funding from the European Research Council (ERC) under the European Union's Horizon Europe research and innovation program (grant agreement No. 101053020, project Dust2Planets). 
D.J.\ is supported by NRC Canada and by an NSERC Discovery Grant.
This paper makes use of the following ALMA data: ADS/JAO.ALMA\#2018.1.01205.L. ALMA is a partnership of ESO (representing its member states), NSF (USA) and NINS (Japan), together with NRC (Canada), MOST and ASIAA (Taiwan), and KASI (Republic of Korea), in cooperation with the Republic of Chile. The Joint ALMA Observatory is operated by ESO, AUI/NRAO and NAOJ. The National Radio Astronomy Observatory is a facility of the U.S. National Science Foundation operated under cooperative agreement by Associated Universities, Inc.
\end{acknowledgements}

\bibliographystyle{aa}
\bibliography{reference}

\onecolumn
\begin{appendix}
\section{ALMA observational epoch}

We present the observational information of three Setup over our observing epoch in Table~\ref{table:a1}, and the continuum imaging summary in Table~\ref{table:cont}.

\begin{table}[!htbp]
\caption{Summary of ALMA observations.}
\centering
\begin{tabular*}{\textwidth}{@{\extracolsep{\fill}}l c c c c c c} 
\hline\hline
Track ID (Freq) & UTC date & Time range & Array (Config.) & Baseline & PWV & Total elapsed time \\
(GHz) & (YYYY-MM-DD) & (h:m:s) &  & (m) & (mm) & (seconds) \\
\hline
\multicolumn{7}{c}{} \\
\multirow{4}{*}{Setup2 (252.95)} & 2018-11-17 & 14:49:22.5-15:49:40.1 & 7m & 7.50-48.65 & 1.800 & 3617.62 \\     
 & 2018-11-29 & 14:01:24.0-14:24:19.8 & 12m (C43-5) & 12.68-1221.06 & 0.663 & 1375.82 \\
 & 2019-01-22 & 14:28:22.2-14:35:27.9 & 12m (C43-2) & 13.97-302.37 & 1.319 & 425.76 \\
 & 2019-04-19 & 06:10:51.8-06:34:28.1 & 12m (C43-4) & 14.64-737.63 & 0.663 & 1416.34 \\
\multicolumn{7}{c}{} \\
\multirow{4}{*}{Setup 1 (225.44)} & 2018-11-14 & 15:25:59.8-16:08:01.5 & 7m & 7.91-48.70 & 1.800 & 2521.68 \\
 & 2018-11-16 & 20:11:29.8-20:53:32.4 & 7m & 5.94-48.95 & 1.800 & 2522.54 \\
 & 2018-11-28 & 14:27:44.4-15:00:15.7 & 12m (C43-5) & 13.35-1229.41 & 1.137 & 1951.39 \\
 & 2019-01-18 & 11:19:32.0-11:30:36.9 & 12m (C43-2) & 13.68-302.86 & 2.271 & 664.90 \\
\multicolumn{7}{c}{} \\
\multirow{2}{*}{Setup 3 (100.62)} & 2018-12-16 & 12:51:41.8-12:56:13.9 & 12m (C43-4) & 12.32-656.17 & 5.289 & 272.16 \\ 
 & 2019-09-24 & 18:11:26.8-18:31:37.2 & 12m (C43-6) & 12.58-2145.75 & 3.056 & 1210.42 \\
\hline
\end{tabular*}

\label{table:a1}

\end{table}

\begin{table*}[htbp]
\caption{Continuum imaging summary.}
\label{table:cont}
\begin{threeparttable} 
\centering
\begin{tabularx}{\textwidth}{@{\extracolsep{\fill}}l c c c c c } 
\hline\hline
\multirow{2}{*}{Track ID (Wavelength)} & RA-Dec (ICRS)\tnote{a} & \multirow{2}{*}{robust} & $\theta_{\rm maj} \times \theta_{\rm min}$ (PA)\tnote{b} & $\sigma_{\rm rms}$\tnote{b} & peak intensity\tnote{b} \\ 
 & 16:mm:ss, -24:mm:ss &  & $^{\prime\prime} \times ^{\prime\prime} (\degree)$ & ($\mu$Jy beam$^{-1}$) & (mJy beam$^{-1}$)  \\
\hline
\multicolumn{6}{c}{IRS 3} \\
\multirow{3}{*}{Setup 1 (1.3\,mm)} & \multirow{3}{*}{26:21.72, 22:51.03} &  2.0 & $0.55 \times 0.44$ (-73.9) & 72.6 & 91.3 \\
 & &  0.5 & $0.46 \times 0.35$ (-65.0) & 62.2 & 78.3 \\
 & & -2.0 & $0.36 \times 0.22$ (-60.2) & 179.0 & 57.5 \\
\multirow{3}{*}{Setup 2 (1.2\,mm)} & \multirow{3}{*}{26:21.72, 22:51.02} &  2.0 & $0.57 \times 0.46$ (-83.7) & 69.7 & 124.9 \\
& &  0.5 & $0.49 \times 0.38$ (-76.5) & 62.7 & 110.5 \\
& & -2.0 & $0.39 \times 0.26$ (-71.4) & 151.1 & 86.7 \\
\multirow{3}{*}{Setup 3 (3.0\,mm)} & \multirow{3}{*}{26:21.72, 22:51.05} &  2.0 & $0.66 \times 0.50$ (80.7) & 69.7 & 20.6 \\
& &  0.5 & $0.52 \times 0.39$ (-82.6) & 67.9 & 18.3 \\
& & -2.0 & $0.38 \times 0.26$ (-67.0) & 249.8 & 14.3 \\
\multicolumn{6}{c}{IRS 2} \\
\multirow{3}{*}{Setup 2 (1.2\,mm)} & \multirow{3}{*}{26:22.39, 22:53.51} &  2.0 & $0.57 \times 0.46$ (-83.7) & 69.7 & 0.55\\     
 & &  0.5 & $0.49 \times 0.38$ (-76.5) & 62.7 & 0.48\\
 & & -2.0 & $0.39 \times 0.26$ (-71.4) & 151.1 & 0.62\\
\multicolumn{6}{c}{IRS 1} \\
\multirow{3}{*}{Setup 1 (1.3\,mm)} & \multirow{3}{*}{26:21.36, 23:04.96} &  2.0 & $0.55 \times 0.44$ (-73.9) & 72.6 & 12.1 \\
 & &  0.5 & $0.46 \times 0.35$ (-65.0) & 62.2 & 11.5\\
 & & -2.0 & $0.36 \times 0.22$ (-60.2) & 179.0 & 10.9\\
\multirow{3}{*}{Setup 2 (1.2\,mm)} & \multirow{3}{*}{26:21.36, 23:04.96} &  2.0 & $0.57 \times 0.46$ (-83.7) & 69.7 & 15.0\\
 & &  0.5 & $0.49 \times 0.38$ (-76.5) & 62.7 & 15.3\\
 & & -2.0 & $0.39 \times 0.26$ (-71.4) & 151.1 & 13.8\\
\multirow{3}{*}{Setup 3 (3.0\,mm)} & \multirow{3}{*}{26:21.36, 23:04.97} &  2.0 & $0.66 \times 0.50$ (80.7) & 69.7 & 3.2 \\
 & &  0.5 & $0.52 \times 0.39$ (-82.6) & 67.9 & 3.1\\
 & & -2.0 & $0.38 \times 0.26$ (-67.0) & 249.8 & 3.0\\

\hline
\end{tabularx}

\begin{tablenotes}
\raggedright
\item[a] Position obtained from two-dimensional Gaussian fits.
\item[b] Values are derived from the primary-beam-corrected image, with $\sigma_{\rm rms}$ measured using a 50\% cutoff.
\end{tablenotes}

\end{threeparttable} 
\end{table*}

\section{UV consistent images}\label{appendix:img}

\begin{figure*}
    \hspace{0.0cm}
    \includegraphics[width=17.5cm]{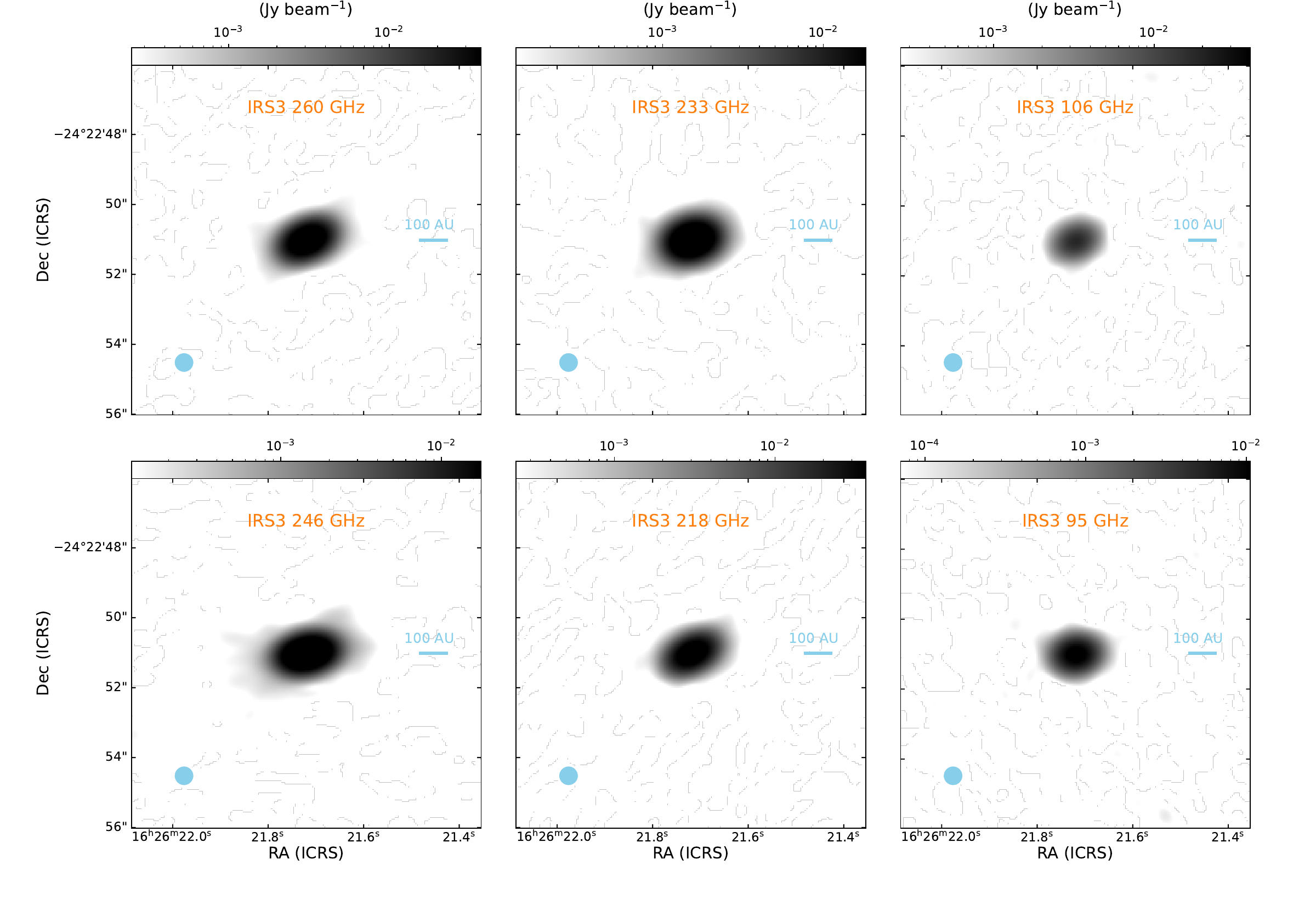}\\
    \includegraphics[width=17.5cm]{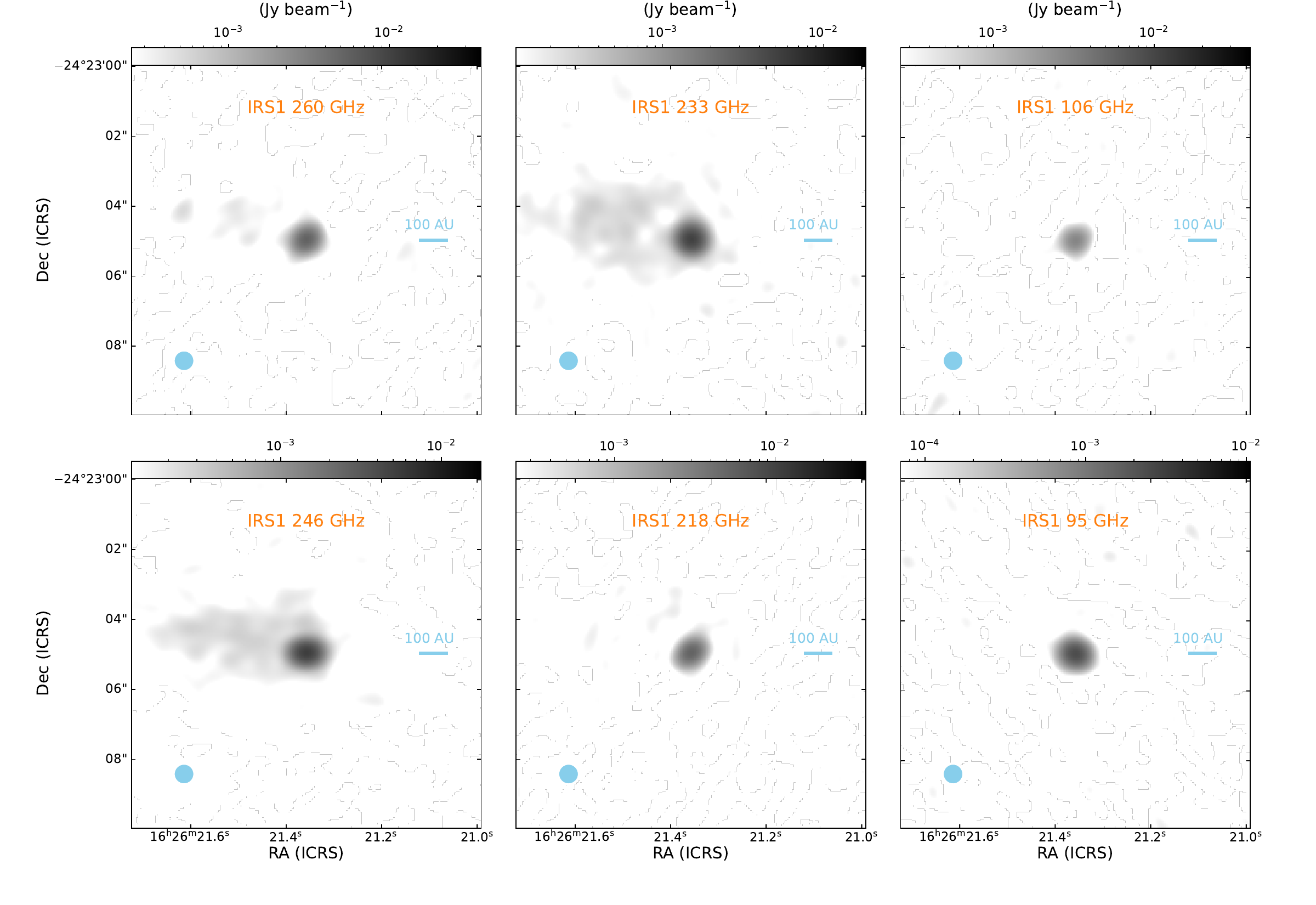}
    \caption{The zoomed-in smoothed sidebands image of IRS3 (top) an IRS1 (bottom).}
    \label{fig:zoomed sideband image}
\end{figure*}

We show the smoothed \textit{uv} consistent sidebands images of three Setup in this section. Figure~\ref{fig:zoomed sideband image} shows the sidebands image of images of IRS3 and IRS1.

\end{appendix}

\end{document}